# On the origin of non-classical ripples in draped graphene sheets

Riju Banerjee[1*#], Tomotaroh Granzier-Nakajima[1], Aditya Lele[2], Jessica A. Schulze[3], Md. Jamil Hossain[2], Wenbo Zhu[2], Lavish Pabbi[1], Malgorzata Kowalik[2], Adri C.T. van Duin[2,3], Mauricio Terrones[1] and E.W. Hudson[1*]

[1]Department of Physics, The Pennsylvania State University, University Park, PA 16802, USA
[2]Department of Mechanical Engineering, The Pennsylvania State University, University Park, PA 16802, USA
[3]Department of Chemistry, The Pennsylvania State University, University Park, PA 16802, USA
[4]Department of Chemical Engineering, The Pennsylvania State University, University Park, PA 16802, USA
[#]Currently at James Franck Institute, University of Chicago, Chicago, IL 60637, USA

*Corresponding authors: rijubanerjee@uchicago.edu, ehudson@psu.edu

**Abstract:** Ever since the discovery of graphene and subsequent explosion of interest in single atom thick materials, studying their mechanical properties has been an active area of research. New length scales often necessitate a rethinking of physical laws, making such studies crucial for understanding and ultimately utilizing novel material properties. Here we report on the investigation of nanoscale periodic ripples in suspended, single layer graphene sheets by scanning tunneling microscopy and atomistic scale simulations. Unlike the sinusoidal ripples found in classical fabrics, we find that graphene forms triangular ripples, where bending is limited to a narrow region on the order of a few unit cell dimensions at the apex of each ripple. This non-classical bending profile results in graphene behaving like a bizarre fabric, which regardless of how it is draped, always buckles at the same angle. Investigating the origin of such non-classical mechanical properties, we find that unlike a thin classical fabric, both in-plane and out-of-plane deformations occur in a graphene sheet. These two modes of deformation compete with each other, resulting in a strain-locked optimal buckling configuration when draped. Electronically, we see that this in-plane deformation generates pseudo electric fields creating a ~3 nm wide pnp heterojunction purely by strain modulation.

**Main text:**

From the folds in our textiles to ripples in plastic wrap and draped tablecloths, deformation of thin sheets are common everyday occurrences. The essential physics of these everyday observations was worked out more than a century ago by Lord Rayleigh in his book *Theory of Sound* [1]. Unlike familiar thin fabrics with finite measurable thicknesses, single atom thick 2D materials are the ultimate limit of thin sheets and understanding how they deform has been the focus of research over the past decade. From a fundamental viewpoint, it is important to understand how mechanical properties of 2D materials, especially rippling, buckling, folding and crumpling, differ from those of classical fabrics. From an application viewpoint, the potential to develop novel electronic devices which exploit the ability of 2D materials to deform into complex three dimensional structures while bending with very small radii makes these studies crucial [2–5]. One of the earliest studies of graphene's mechanical properties was done by Lee et al. [6] by making indentations in suspended graphene with an atomic force microscope. They found that graphene has a Young's modulus of 1TPa (five times that of most steels) and an intrinsic strength of 130 GPa (fifty to a hundred times that of most steels) – the highest for any material ever measured. On the theoretical side, efforts were also made to go beyond numerically expensive first principle calculations towards equivalent continuum models that can explain the mechanics of larger graphene sheets as a fabric [7–9]. Finding an equivalent description of 2D materials analogous to classical fabrics is challenging for two main reasons. First, the inability to properly define the plate thickness of a single atomic layer results in wide variations in bending and Young's moduli among different studies, known as Yakobson's paradox [10,11]. Secondly, an equivalent description of a discrete lattice is challenging when the length scale of strain variation becomes comparable to the lattice constant. In particular, it is plausible that Rayleigh's theories get modified when deformation length scales approach the atomic limit, when electronic interactions (which were unknown when the classical theory was developed) between neighboring atoms can become important. Indeed, Tapaszto et al. [12] found that nanometer length scale graphene ripples have shapes that cannot be described by a continuum mechanics approach valid for a classical fabric [13]. Interestingly, Bai et al. [14] found this discrepancy persists for ripples measuring up to 100 nm, length scales much larger than the lattice constant, whereas micron sized ripples were well described by continuum mechanics [15]. This suggests two regimes in which ripples in graphene follow different sets of laws. Continuum mechanics describes well graphene ripples with a wavelength of around one micron and greater, and another, unknown set of laws governs rippling of graphene on the length scales below 100 nm. Here we investigate the origin of the latter,with graphene ripple wavelengths on the order of tens of nanometers.

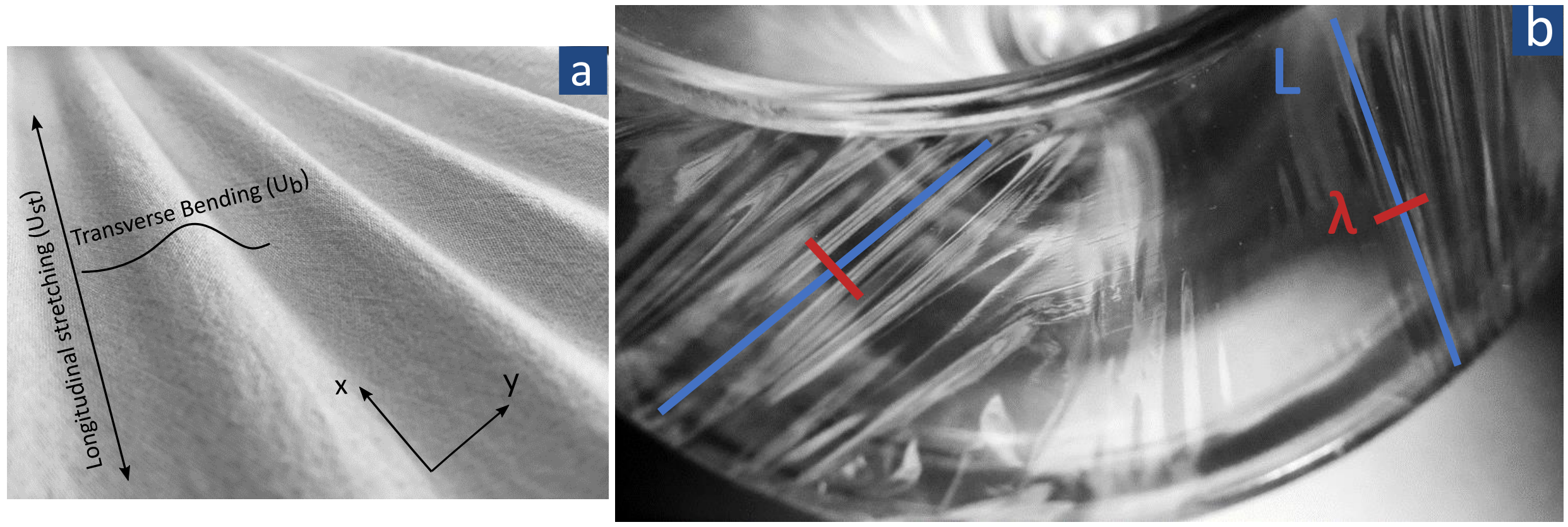

**FIG 1: Rippling in classical fabrics:** (a) Applying tensile stress to a fabric in the longitudinal direction (x) results in a Poisson compression in the transverse direction (y), creating ripples. The shape of ripples in a classical fabric is determined by balancing energy contributions from two deformations: stretching in the longitudinal direction (with corresponding energy cost $U_{st}$) and bending in the transverse direction (energy cost $U_b$). (b) Plastic wrap spread over two concentric cylinders, with taller inner cylinder, also yields such tension ripples. As strain $\gamma$ is kept constant, ripple wavelength $\lambda$ and sheet length $L$ follow a scaling law $L \propto \lambda^2$, as in Equation 2 of the text. In the figure, ripples on the right have smaller wavelength (red) and amplitude than those on the left. But to stabilize the larger ripples, the left ripples must run diagonally, increasing the effective sheet length $L$ (blue).

Ripples are necessary for the stability of 2D materials [16–18], and are commonly observed in scanning tunneling microscopy (STM) investigations of graphene. However, development of a thorough formalism of 2D material deformation at the nanoscale, equivalent to that for classical fabrics, has been hindered due to the difficulty of creating reproducible nanoscale ripples such that physically relevant parameters like wavelength, amplitude, sheet length and curvature can be systematically varied and measured over a large parameter space. Instead, experimental studies have typically relied on the good fortune of finding wrinkles in 2D materials [14,19–24]. Though it is possible to induce controlled strain by patterning substrates [25–27], this approach tends to create wrinkles with deformation length scales much larger than the lattice spacing. In contrast to these traditional approaches, we develop a technique to create reproducible nanoscale ripples by draping graphene over Cu step edges and experimentally investigate a large parameter space of wavelengths, amplitudes and sheet lengths.

In describing the deformation of a classical sheet, Rayleigh argued [1] that there are two ways a uniform isotropic thin sheet can deform: by out of plane bending and by in-plane stretching. When a sheet of thickness $t$ bends with a radius of curvature *R*, then the strain due to bending, $\gamma_b$ , varies through its thickness as $\gamma_b \sim z/R$ where *z* is the axis normal to the sheet surface. Hence the energy density due to bending, $U_b$, is $U_b \sim \int {\gamma_b}^2 \ dz \sim t^3$. Similarly, the energy density due to a uniform stretch is $U_{st} \sim \int {\gamma_{st}}^2 \ dz \sim t$. Thus the total energy *U* due to deformation is

$$U \sim t^3(\text{Bending}) + t(\text{Stretching}). \qquad (1)$$

As the deformation should minimize energy cost, Rayleigh remarked 'when the thickness is diminished without limit, the actual displacement will be one of pure bending' [1]. Scaling laws relating wavelength, sheet length and strain magnitude for a thin classical rippled sheet were deduced by Cerda and Mahadevan [13,28]. In particular, they showed that ripples in a classical fabric (Fig. 1a) obeys the scaling law relating wavelength $\lambda$, strain $\gamma$, sheet length $L$ and thickness $t$

$$L \sim \frac{\sqrt{\gamma}\lambda^2}{t} \qquad (2)$$

This scaling law implies that for a sheet under constant strain, the sheet length *L* varies as $\lambda^2$. An example of this is illustrated in Fig. 1b, where the ripples on the left have a larger wavelength and amplitude than those on the right, but to accommodate the larger ripples, the sheet must drape down diagonally, effectively increasing the sheet length. We sketch the origin of this scaling law in supplementary section 9.

For most STM studies of graphene an atomically flat substrate is desired in order to facilitate the growth of large grains and to ease scanning. However, to encourage the rippling of graphene in our study, we grow graphene via low pressure chemical vapor deposition at high temperature (1020°C) on an electropolished Cu substrate (see supplementary section 1). Under these growth conditions, step bunching leads to the formation of large steps up to 35 nm tall [29,30] (Fig. 2a). Graphene drapes down these large steps forming ripples (Fig. 2b) in the draped(suspended) regions.

These ripples  occur preferentially in regions of higher step edge curvature (Fig. 2c, d). This is analogous to the classical case of ripples forming at the corners of a tablecloth, a result of having more expendable material at that location.  In Fig. 3 we present measurements of their wavelength, amplitude, and sheet length. The wide variation of parameters from more than two hundred independent measurements across multiple steps enables us to explore a large parameter space. This, in turn, allows us to avoid the necessity of estimating experimentally unmeasured quantities like the effective thickness of the graphene sheet and its Poisson's ratio at the nanoscale [12,14], and instead look for scaling relationships to directly test classical laws (Figure 3).

Fig. 3c shows that contrary to a sinusoidal profile of ripples observed in classical fabrics, these ripples have a triangular shape. More interestingly, all ripples have the same bending angle at the vertex, $168^o$ (±3°), consistent across order of magnitude variations in wavelength (Fig. 3d), amplitude, and step height. This conserved bending angle means that the ripple's wavelength is proportional to its amplitude. Such a situation arises in the classical theory for ripples with uniform strain due to tranverse inextensibility. A sheet that is stretched in the longitudinal direction with strain $\gamma$ Poisson contracts in the transverse direction. Transverse inextensibility of the sheet implies that when a flat region of width $\lambda'$ buckles forming a ripple of wavelength $\lambda$ and amplitude A, then they are related by $\left(\frac{\lambda'}{2}\right)^2 \approx \left(\frac{\lambda}{2}\right)^2 + A^2$. The contraction in the transverse direction due to this out of plane buckling is $\Delta\lambda = \lambda' - \lambda \approx \frac{2A^2}{\lambda}$ for $\Delta\lambda \ll \lambda$. As this contraction results in the Poisson compression, $\gamma_{transverse} = \nu\gamma = \frac{\Delta\lambda}{\lambda} = \frac{2A^2}{\lambda^2}$, where $\nu$ is the Poisson's ratio. Thus, we have $A \propto \sqrt{\gamma}\,\lambda$, as also derived in ref. [13].

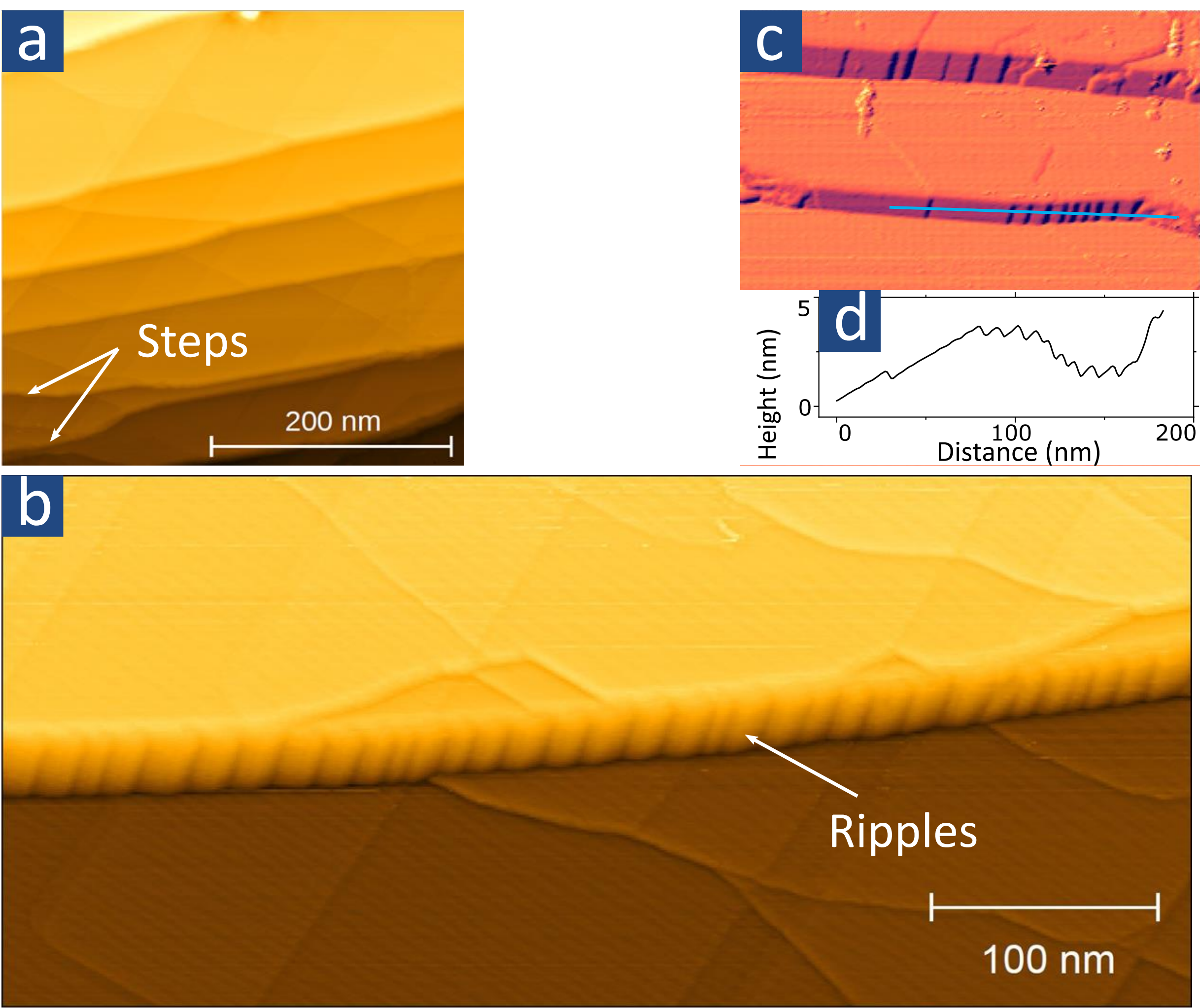

**FIG 2: Creating nanoscale ripples in graphene:** (a) STM topograph of graphene grown on copper with large (up to ~35 nm high) step edges (current setpoint $I_{set}$ = 70 pA, sample bias $V_s$ = 0.1 V; all displayed STM data in the paper is obtained at temperature T = 80 K and unfiltered). (b) Magnified image of a step edge ($I_{set}$ = 80pA, $V_s$ = 0.1 V) shows graphene draping over the step forming ripples as it is stretched by the contact forces of the Cu substrate. (c) Analogous to corners of a draped tablecloth, high curvature regions are associated with higher ripple density (and hence smaller ripple wavelength) as highlighted in topographic profile (d), extracted along the blue line ($I_{set}$ = 170 pA, $V_s$ = 0.12 V).

These two scaling laws, viz., $A \propto \sqrt{\gamma}\,\lambda$ and $L \sim \frac{\sqrt{\gamma}\lambda^2}{t}$ (equation 2) implies that for any two arbitrary ripples in a classical fabric, only two scenarios are possible: either (i) $\gamma$ is same, in which case both $A\sim \lambda$ and $L\sim\lambda^2$; or (ii) $\gamma$ is different, in which case both $A \nsim \lambda$ and $L \nsim \lambda^2$. As we observe a conserved bending angle at the crest of each ripple (Fig. 3d), the ripples satisfy the scaling relation $A\sim \lambda$. If the observed ripples are classical, then from case (i) above, they must also satisfy $L\sim\lambda^2$. We test this prediction directly by measuring sheet length $L$ and wavelength $\lambda$ of ripples. Fig. 3a displays two regions on the same step where measurements were taken. For these two regions, the local ripple wavelengths, shown in Fig. 3c along the ripple direction $\mathcal{A}_2$ and $\mathcal{B}_2$, differ by a factor of 2 ($\lambda_B \approx 2\lambda_A$) while the sheet lengths (Fig.3b) are nearly identical, $L_B \approx L_A$ in strong contrast with the classical prediction of $L_B \approx 4L_A$ (from Equation 2). A plot of sheet length versus wavelength for several ripples observed on three different step edges (Fig. 3e) highlights the lack of any clear correlation between the two variables. The disagreement between scaling laws predicted by the classical theory of rippling and our measurements demonstrate that classical theory is not valid for nanoscale ripples in graphene. We also note about a potential caveat that the classical scaling law $L \propto \lambda^2$ is strictly valid for a flat geometry, but the ripples were observed on curved step edges, and whether that can invalidate our arguments showing graphene to be non-classical. We address this issue in supplementary section 9, where we show that as the local radius of curvature of the underlying step edges are several orders of magnitude larger than all the other length scales in the problem, like wavelength, amplitude and sheet length, it only adds negligibly small higher order corrections to the classical scaling law.

We next turn to addressing some potential issues that can arise when imaging 2D materials on a tilted surface using STM. Tip artifacts are common when imaging 2D materials with STM, particularly because they tend to stick to the tip. However, such effects are generally more pronounced for unstretched micron sized suspended sheets [25]. The graphene sheet in our study should be much less susceptible to such behavior as it is under tension and is suspended only over a very small region (tens of nanometers). Topographs taken in forward and backward scans, as well as under order of magnitude ranges of tunneling current (~100 pA to 1 nA) are similar, implying that graphene is not sticking to the tip. The experimentally measured rippling angle and its triangular shape (Fig. 3d) is also found consistent with atomistic simulations (presented later in Fig. 4h) further supporting that tip artifacts are negligible in the experimental data. Imaging a tilted region, like the draped region, can also create artifacts as the standard technique of plane subtraction can lead to artificial compression. Instead, we rotate the coordinate system to extract accurate lengths on draped graphene (supplementary section 7).

To understand the origin of these non-classical mechanical properties, we performed atomistic scale calculations of a graphene sheet draped over a curved Cu step edge (Figure 4). While density functional theory (DFT) calculations are widely used to model graphene structures [31–35], such simulations are computationally expensive, and not feasible for modeling a graphene sheet suspended over a Cu step edge, which requires consideration of at least a few thousand atoms. On the other hand, none of the widely used classical force fields - which are generally suitable for simulating large systems - such as AMBER [36], CHARMM [37], GROMOS [38] or OPLS [39], can be used to correctly reproduce the mechanical properties of graphene while simultaneously satisfying chemically justified carbon-copper interactions. Fortunately, a recently developed Cu-C ReaxFF parameter set [40] can be used to model a graphene sheet at a copper step edge, as it not only includes the parameters for interactions between carbon atoms which correctly describe the mechanical properties of graphene, but has also been shown to accurately model experimentally observed draping of a suspended graphene ribbon at a copper step [41]. Thus, we used the ReaxFF method with a recently updated version of this parameter set to further stabilize the carbon-copper interactions to perform atomistic simulations and understand the mechanical properties of our system. See supplementary section 3 for more information on the ReaxFF simulations performed. All presented simulations snapshots were visualized using VMD (Visual Molecular Dynamics) [42] or the Schrodinger package [43], while all atomistic simulations were performed with use of the ADF software [44].

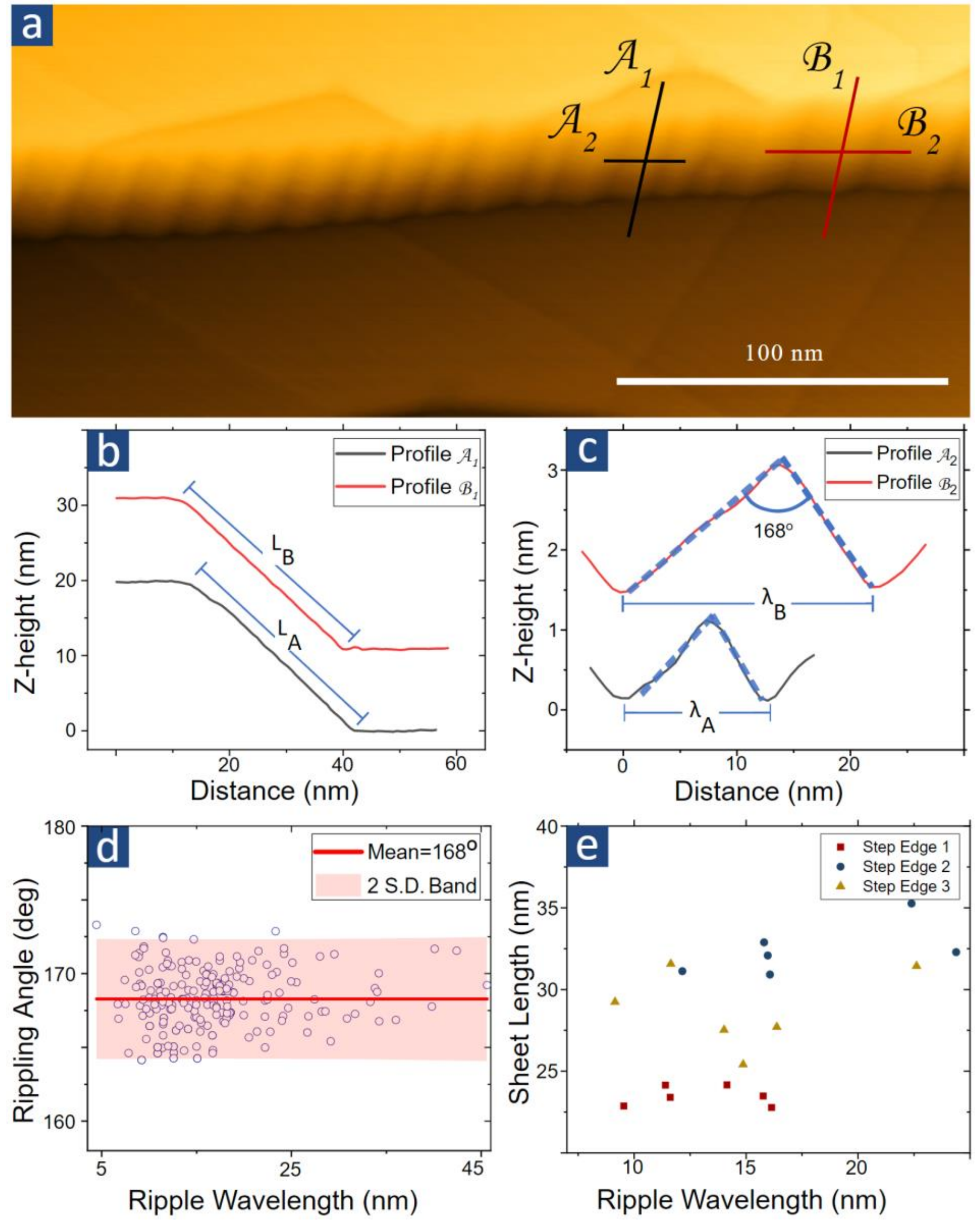

**FIG 3: The shape of graphene:** (a) Topography of ripples on draped graphene ($I_{set}$ = 30 pA, $V_s$ = 0.1 V) from whence we extract profiles (b) down the sheet ($\mathcal{A}_1$/$\mathcal{B}_1$), revealing sheet length and (c) along the sheet ($\mathcal{A}_2$/$\mathcal{B}_2$), showing ripple wavelength and apex angle. Dashed lines highlight the triangular shape. Profiles are offset for clarity and have different scales for the horizontal and vertical axes. (d) The rippling angle is conserved at $168°(\pm 3°)$, independent of step height and curvature, and ripple amplitude and wavelength (pictured here, with two standard deviation range shaded pink). (e) Measurements of sheet length and ripple wavelength from three separate step edges show the violation of classical scaling, which predicts $L \propto \lambda^2$.

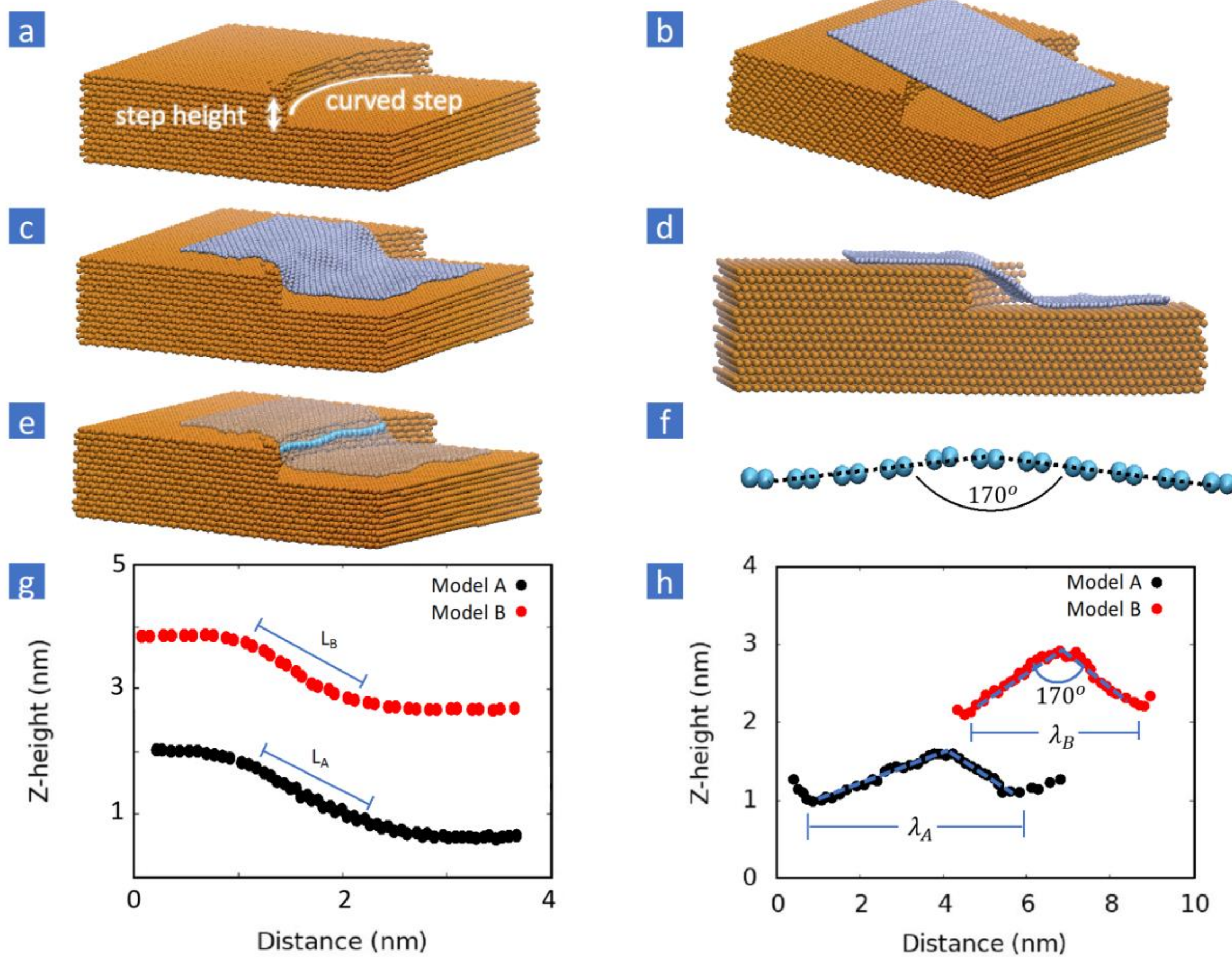

**FIG 4: Atomistic scale simulations:** Building process and simulation results for an atomistic representation of a curved Cu step with rippled, suspended graphene. To investigate formation of ripples in graphene draped over a curved Cu step edge (a), we place a sheet of graphene in close proximity to this Cu step (b) and let the system relax to form rippled, suspended graphene (c). A cut along the crest of the ripple, showing only half the system is presented in (d), while the ripple profile can be seen in the highlighted cyan atom chain (e) and in an extracted profile showing the apex angle (f). Profiles obtained from the final structure for two models with different Cu-step radii of curvature (Model A: 10nm; Model B: 15nm; see supplementary section 3) (g) down the sheet reveal sheet length and (h) along the sheet show ripple wavelength and apex angle. The profiles are offset for clarity.

As the ripples are observed to form preferentially on curved step edges (Fig. 2c, d), we model them by generating a series of Cu steps of varying height *h* and curvature (Fig. 4a) and placed 8 nm square graphene sheets near them (Fig. 4b). All considered structures were first energy minimized, proceeded by a seriesof NPT (i.e., keeping the number of particles, the pressure and temperatures constant) simulations. The sheet temperature was initially set to 5 K to allow the graphene to slowly connect with the Cu substrate. Then, to facilitate deformations of the graphene sheet including additional attachment to the substrate, the sheet temperature was raised to 175 K. After 15 ps the graphene formed visible ripples for all considered systems. In Fig. 4c the final structure for one such model (described as model A in supplementary section 3) with step height 1.6 nm is presented. Unlike the STM data, where it is not possible to see underneath the graphene sheet, for the simulated system we can create a cut-away view (Fig. 4d), highlighting suspension of the graphene in the draped region. Thus, the simulation implies that ripples similar to those experimentally observed can be induced in graphene just by draping it on a curved Cu step edge (analogous to a draped tablecloth) and does not require the formation of any vicinal surface reconstructions in the substrate which might occur at high growth temperatures [45].

Next, we compare the shape of the ripples obtained from these simulations and find them to be consistent with experimental observations. To better visualize the shape of the rippled graphene sheet, in Fig 4e we make all carbon atoms translucent except for a line of atoms across the ripple highlighted in cyan. In the extracted profile of these highlighted atoms (Fig. 4f) we see that these simulated ripples are also triangular (as highlighted by black dashed linear guides to the eye) rather than sinusoidal, with an apex angle of $170°$ ($\pm 2°$) across all considered systems (Table T2 in supplementary section 3), consistent with the experimentally measured value of $168°$ ($\pm 3°$). Furthermore, profiles extracted from two different simulated systems show that it is possible to have ripples with similar sheet lengths (Fig. 4g) but very different ripple wavelengths (Fig. 4h - here $\lambda_A \approx 1.5\lambda_B$). We also tested the system with a rotated graphene sheet to comfirm that chirality does not play a significant role in determining the ripple shapes (Fig. S2 in supplementary section 3). Thus, our simulations accurately describe the observed mechanical properties of these nanometer wavelength ripples in draped graphene.

With simulations capturing the essential physics of the system, we next try to understand these remarkable observations. In particular, we try to understand exactly which parts of classical rippling theory fail when applied to a graphene sheet and why. One of the advantages of our simulation approach is the possiblility of selectively switching off particular interactions to assess how important they are for a given phenomena to occur. As elaborated further in supplementary section 3, we find that switching off dihedral interations immediately leads to the disappearance of these ripples (Fig. S2). The dihedral interaction, which is necessary to correctly describe the presence of conjugated $\pi$ bonds, captures the effect of bending of a discrete lattice and hence it is only natural that it plays a crucial role in determining the shape of the ripples. Lu et al. [46] found that the dihedral interaction is important in determining the bending modulus of monolayer graphene. In particular, they hypothesized using theoretical calculations that carbon nanotubes must stretch in the tangential direction when it is rolled from a graphene sheet, increasing its radius.

These two observations, that ripples are eliminated when dihedral interactions, are turned off and that dihedral interactions can result in transverse stretching of the bonds, suggest that this transverse in-plane stretching is likely a crucial component missing from the classical theory of rippling. The classical theory only considers out of plane bending, and that may lead to incorrect predictions when applied to ripple structures in graphene (and likely other 2D materials). Moreover, such in-plane stretching, should it exist, would likely occur only where graphene is bent, that is, as the ripples are triangular, in a very narrow region at the crest of each ripple. Such a strain variation over a very narrow region is challenging to resolve using direct atomic imaging techniques. For example, though such an in-plane mode of deformation was also proposed by Tapaszto et. al. [12] to account for the breakdown of the classical rippling theory, direct experimental evidence of the same and measurement of this in-plane strain and its spatial variation has not been achieved. For that, we turn to measuring electronic properties which should be affected by this local stretching to provide indirect evidence of in-plane stretching of bonds at the ripple extrema. Electronic properties are dependent on the overlap of electronic wavefunctions of nearest neighbor atoms, and are hence exponentially more sensitive than STM topography to bond length distortions.

In Fig. 5a, we show the local density of states (LDOS) measured on rippled graphene, where, in addition to the V-shape of the Dirac cone, we see a series of equidistant peaks due to the formation of strain modulated superlattices [47]. The peaks are most prominent for negative biases, with the asymmetry arising likely due to next nearest neighbor hopping in a strained lattice [48]. We focus only on those bias values for our next analysis. Plotting the LDOS at each point of a linecut profile near a ripple crest (Fig. 5b, c) we see that the equidistant peaks (marked by red dashed lines) shift in unison towards negative bias, implying a shift of the Dirac point (DP). Such a DP shift is known to occur due to changing of the local scalar potential and reduced Fermi velocity when graphene is stretched [49–52]. Following Guinea et al. [53], we use this DP shift to estimate strain at the ripple crest. A one dimensional stretching of the lattice $u_{xx}$ creates a scalar potential $V = V_0\, u_{xx}$, where $V_0 \approx 3\ eV$. A DP shift of roughly 100 meV over 3 nm implies a pseudo-electric field of strength $|E| = \frac{100\ meV}{3\ nm} = 3 \times 10^7\,\frac{\mathrm{V}}{\mathrm{m}}$, suggesting a transverse stretching of about $u_{xx} = \frac{|E|x}{V_0} = 3\%$. We also note that this pseudo-electric field magnitude is comparable to E-fields in the smallest operational p-n junctions [54,55].

A possible origin for this transverse in-plane stretching is explained in Fig. 5d, e. Graphene consists of $sp^2$ bonded carbon atoms with the $p_z$ orbitals sticking out of plane. For a flat graphene sheet, these $p_z$ orbitals have minimal overlap. However, a bent graphene sheet can force these $p_z$ orbitals to interact. By local in-plane stretching of the bonds, the distance between the $p_z$ orbitals can be increased, and hence graphene, unlike a thin classical fabric, suffers in-plane stretching when it is subjected to out of plane buckling. The local charge inhomogeneity resulting from such bending and the subsequent re-hybridization of electronic orbitals should also help in stabilizing the ripples [56]. Numerous theoretical studies have treated in-plane and out of plane deformations of graphene to be independent, analogous to classical fabrics [57–59]. Our results suggest that such an approach might be questionable at the nanometer length scale as these two modes of deformation are inherently tied together even for small out of plane deviations (~$10°$). In addition to the above analysis, based on rigorous atomistic calculations, in supplementary section 5 we also present a simple phenomenological model to capture the mechanics of graphene bending. There, we show that the in plane and out of plane modes of deformations compete with each other resulting in a strain-locked optimal buckling configuration.

We note that the ReaxFF calculations were performed for much smaller model systems than those observed experimentally. The largest simulated system had ~50000 atoms with a Cu step edge height of only 1.6 nm, resulting in fewer and smaller ripples than those observed experimentally. Nevertheless, the fact that similar observations were made as in the experiments suggests that the essential physics of the system is adequately described even by such small systems. Furthermore, in the simulations, the graphene was placed near the curved Cu step edge, meaning that the dynamics of the growth process was neglected, which might also affect the final draping configuration. Large scale ReaxFF simulations, using force-biased Monte Carlo, as previously employed for carbon-nanotube growth in Ni-clusters [60], may be able to capture these growth dynamics. Finally, our observations also suggest that electronic interactions can affect the mechanical properties, forcing graphene to stretch at the ripple extrema (Fig. 5). Electronic interactions, like electron transfer between Cu and graphene, are not explicitly accounted for in the ReaxFF method, but recent reports indicate [61] that perhaps e-ReaxFF might be able to include such effects.

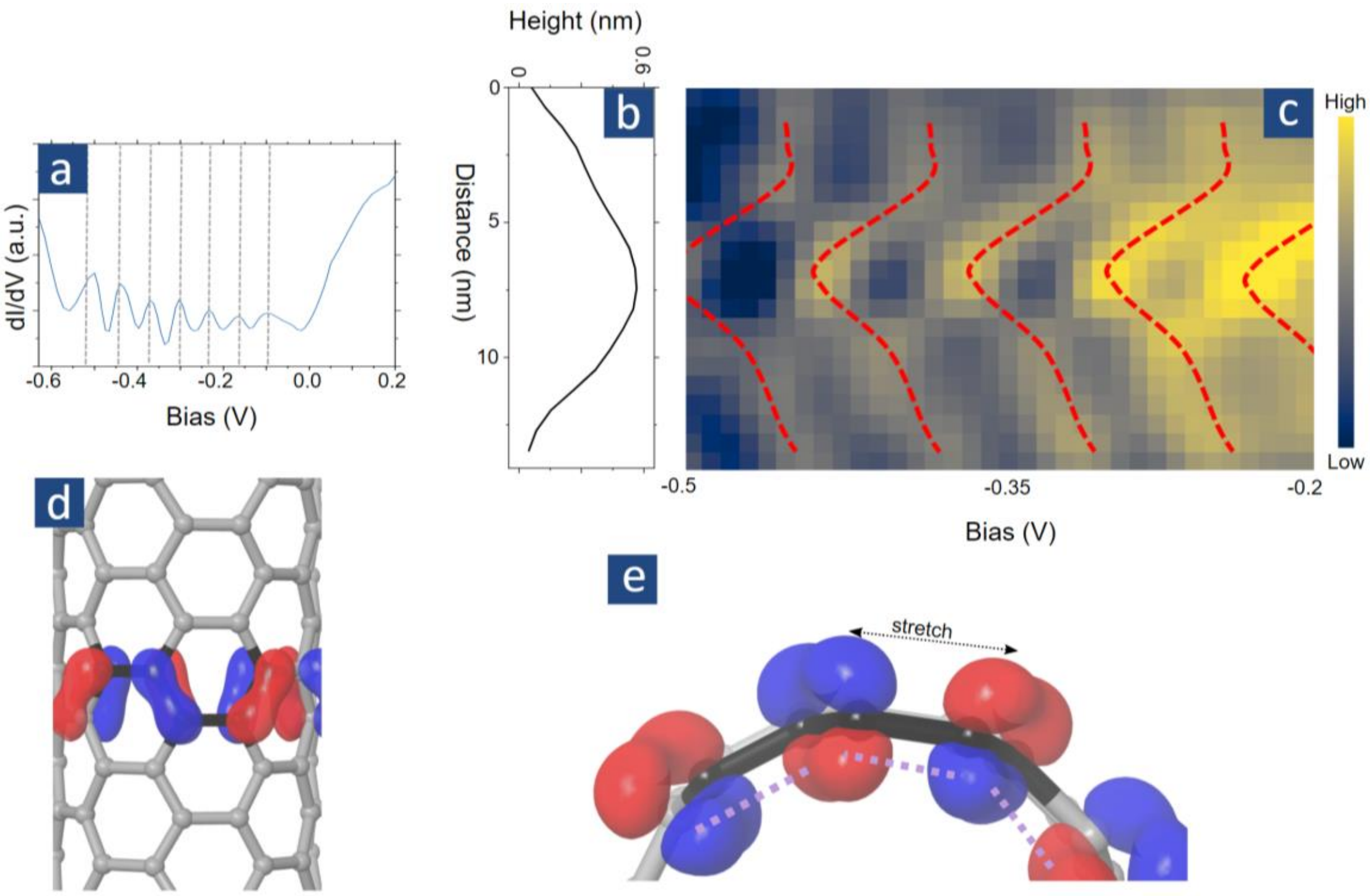

**FIG 5: Origin of non-classical ripples:** (a) Measuring the LDOS by scanning tunneling spectra taken on rippled graphene show a series of equidistant peaks marked by grey dashed lines due to the formation of strain modulated superlattices [47]. Near ripple extrema with height profile shown in (b), the equidistant LDOS peaks shift towards negative bias as marked by red dashed lines in (c). Such a simultaneous shift of all the peaks is the result of shifting of the Dirac point, corroborating that graphene is indeed stretched at the ripple crest. For plot in (c), 100 individual spectra were taken at each spatial point and averaged. When the graphene lattice is bent under curvature (d) top and (e) side views, the $p_z$ orbitals (shown by red and blue lobes with the colors corresponding to electronic spin) get closer than what they were in a flat graphene sheet, forcing them to interact (purple dashed lines). Then, the optimal configuration of graphene is where the local bond lengths are stretched to reduce $p_z$-$p_z$ orbital interaction. This local stretching of the bonds at ripple crests makes it behave unlike a thin classical fabric. These out of plane and in plane modes of deformation occur simultaneously to create a strain-locked optimal buckling configuration in draped grapehen sheets.

We studied graphene sheets as they drape over Cu step edges and explained the formation of ripples that violate classical scaling laws. Unlike a thin classical sheet, we found that graphene undergoes stretching over a few bond lengths at ripple crests and troughs to reduce interactions between nearby $p_z$ orbitals. Interestingly, the effect of this stretching over a very narrow region is significant enough that even ripples much larger than unit cell dimensions (up to tens of nanometers, where one might be tempted to use a continuum approximation) differ fundamentally from classical fabrics. This observation should be considered carefully while developing strain enabled and flexible electronics. Furthermore, the general nature of our arguments imply that other 2D materials should also form non-classical ripples when forced to buckle at the nanoscale. Suggestively, we note that triangular ripples have already been observed in $WSe_2$ as well [62]. In contrast to our study of ripples with wavelengths measuring tens of nanometers, larger ripples, with wavelengths approaching the micron scale, have been shown to conform to the classical theory [15], implying the existence of a crossover region where the bending mechanics of graphene should vary significantly. Such a variation, where the graphene sheet bends differently depending on the size of deformations, may help address the long-standing debate about the bending modulus of monolayer graphene [3,63–65]. Also interesting would be the exploration of how the hybridized $p_z$ orbitals at the ripple crests can affect the local interlayer slipping and ultimately bending in multilayer graphene sheets [66]. The techniques developed in this paper to exert strain at the nanoscale and accurately modeling the effects of such should present a strong foundation for future work in understanding the mechanical properties of 2D materials and realizing the promise of straintronics.

**Funding:** This material is based upon work supported by the National Science Foundation under Grant No. 1229138. T.GN and M.T. acknowledge The Air Force Office of Scientific Research (AFOSR) grant 17RT0244. Computations for this research were performed on the Pennsylvania State University's Institute for Computational and Data Sciences' Roar supercomputer. This content is solely the responsibility of the authors and does not necessarily represent the views of the Institute for Computational and Data Sciences.

**Author contributions:** R.B. conceived the project; T.GN. prepared the samples; R.B.,L.P. built the custom instrument; R.B., L.P. collected the data; R.B., E.W.H. analyzed the data; M.K., A.L., J.A.S., J.H., W.Z., performed the atomistic simulations; R.B., T.GN. did the phenomenological modeling, all authors took part in interpreting the results and writing the paper; A.C.T.vD, M.T., E.W.H. advised.

**Competing interests:** Authors declare no competing interests

# Supplementary Material

## On the origin of non-classical ripples in draped graphene sheets

Riju Banerjee[1*#], Tomotaroh Granzier-Nakajima[1], Aditya Lele[2], Jessica A. Schulze[3], Md. Jamil Hossain[2], Wenbo Zhu[2], Lavish Pabbi[1], Malgorzata Kowalik[2], Adri C.T. van Duin[2,3], Mauricio Terrones[1] and E.W. Hudson[1*]

[1]Department of Physics, The Pennsylvania State University, University Park, PA 16802, USA
[2]Department of Mechanical Engineering, The Pennsylvania State University, University Park, PA 16802, USA
[3]Department of Chemistry, The Pennsylvania State University, University Park, PA 16802, USA
[4]Department of Chemical Engineering, The Pennsylvania State University, University Park, PA 16802, USA
[#]Currently at James Franck Institute, University of Chicago, Chicago, IL 60637, USA

*Corresponding authors: rijubanerjee@uchicago.edu, ehudson@psu.edu

# 1. Growth and Raman characterization:

In preparation for growth, a copper foil is electropolished to smooth its surface. Growth is performed in a low-pressure quartz tube furnace. A piece of electropolished copper is placed in a quartz tube and positioned in the center of a high temperature furnace and a boat containing ammonia-borane is placed in the tube upstream of the furnace and wrapped with a heating belt. Subsequently the tube is pumped down to $10^{-2}$ Torr and flushed with 37.5 sccm $H_2$ and 212.5 sccm Ar gas for several minutes. Afterwards the gas is left flowing, and the furnace temperature is raised to 1020°C at which point the heating belt is set for 50 °C. When the heating belt reaches the set temperature, 10 sccm methane is flowed through the furnace for 5 min. Afterwards the methane is shut off and the furnace cooled naturally to room temperature. At these high growth temperatures, the Cu substrate can diffuse and merge to form large step edges [1]. Presumably, the ripples seen on draped graphene result from the diffusive motion of the Cu steps at the high growth temperatures as graphene is pinned to the Cu substrate by the Van der Waals forces. Some boron and nitrogen dopant atoms were seen embedded in the graphene lattice. They show up clearly as bright spots with STM due to their higher density of states. None were seen in regions where the strained graphene was observed. Raman spectroscopy (Fig. S1) shows that the graphene is indeed monolayer with a characteristic 2D peak. No signature of strain was seen in the Raman spectra, probably because the spot size of several microns was much larger than the width of the draped regions (tens of nanometers).

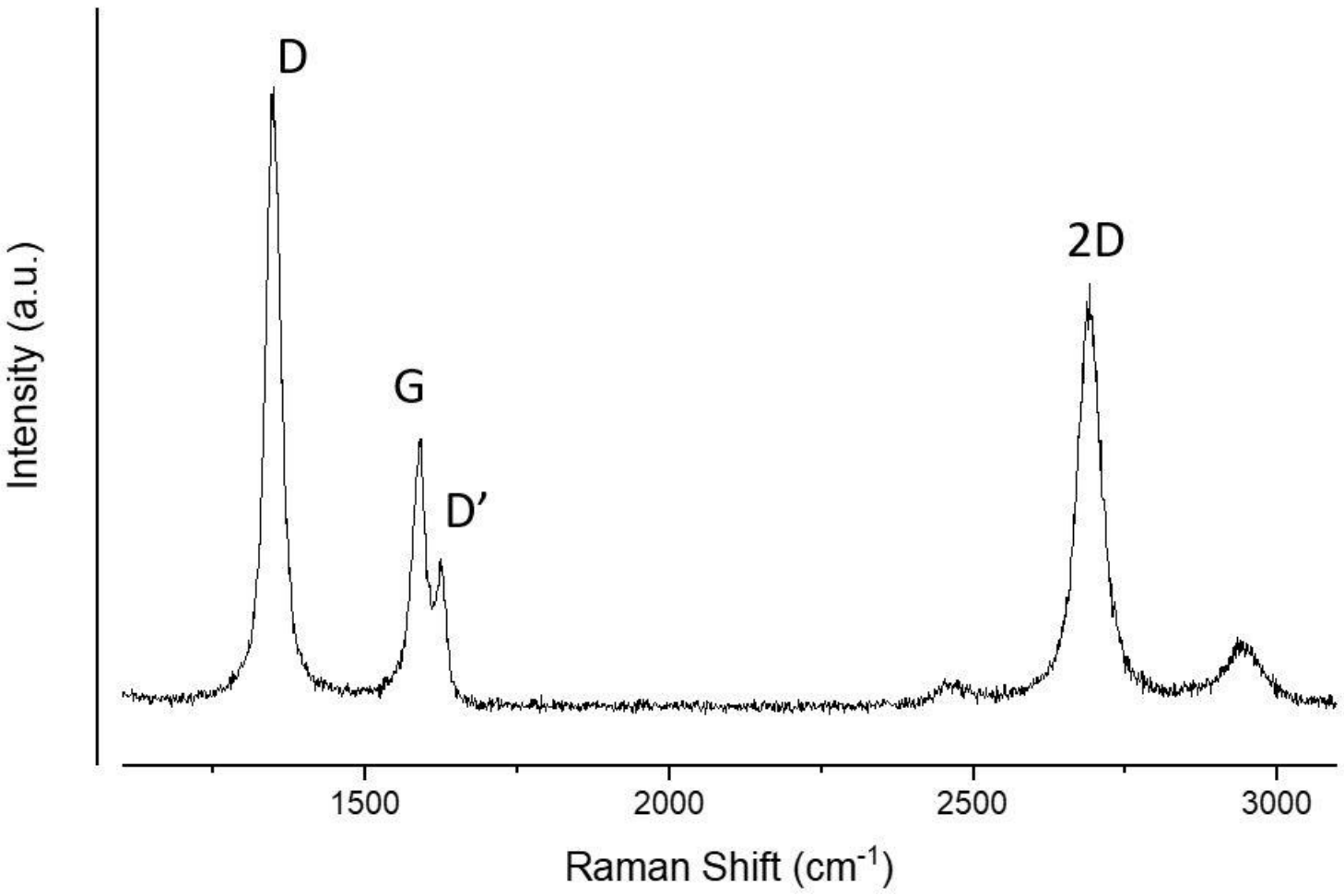

**Figure S1:** Raman spectrum shows the 2D nature of the graphene sheet [2,3].

## 2. Experimental methods

All data was obtained at 80 K in a custom built ultra-high vacuum (UHV) STM system using a SPECS Tyto head with cut Pt-Ir (80%-20%) tips. Part of the analysis was done using the software Gwyddion [4]. Though similar results have been observed with multiple tips on multiple samples, for consistency and calibration, all results presented here are obtained with a single tip on the same sample. Samples were transferred to the UHV environment within 10 minutes of growth to minimize air exposure. The sample was annealed at 300° C for about 1.5 hours in UHV to evaporate any adsorbent that might have settled on the surface during the transfer process. Similar observations have been made even after multiple annealing processes.

## 3. ReaxFF method

ReaxFF is an empirical reactive method where the bond order for all atoms is updated every step during the course of the simulation. The values for all bond orders are determined based on the interatomic distances and all partial energies are bond order dependent. More detailed description of the ReaxFF method can be found e.g. in the review by Senftle et al. [5].

The ReaxFF parameter set previously published by Zhu et al. [6] was further optimized for carbon interactions at various copper surfaces. To obtain a reference for the Cu/C parameters optimization, a series of DFT calculations were performed in the Vienna Ab initio Simulation Package (VASP) [7]. The basis set used was PAW with PBE functionals and a cutoff energy set to 550 eV. A carbon atom was placed at different positions on a Cu slab and the corresponding energy of the configuration calculated. These differences in carbon atoms binding to various subsurfaces are considered in the updated ReaxFF parameter set used for all atomistic simulations reported in the current work.

To systematically study the effect of graphene draping over the Cu step edges, we simulated several different scenarios with different Cu step heights (0.8 nm and 1.6 nm) and radii of curvature (10 nm, 15 nm and 20 nm) (see Table T1). For all cases, 8 nm square graphene sheets were draped over the step edge, except for the model with the step height of 0.8 nm and curvature of 20 nm, where a 19 nm graphene sheet was used. The smallest considered system consisted of 29,158 atoms, while the largest system consisted of 50,512 atoms.

For all simulations we used periodic boundary conditions, a time step of 0.1 fs and a Berendsen thermostat with a damping parameter of 1000 fs. In the case of constant pressure simulations, a Berendsen barostat was used with a damping parameter of 1000 fs. The final characteristics of all considered models are summarized in Table T2.

| Characteristic | Model A | Model B | Model C | Model D | Model E |
| --- | --- | --- | --- | --- | --- |
| Cu step radius of curvature | 10 nm | 15 nm | 10 nm | 15 nm | 20 nm |
| Length of graphene | 8 nm | 8 nm | 8 nm | 8 nm | 19 nm |
| Number of atoms | 37119 | 37913 | 29158 | 30453 | 50512 |
| Z-height | 1.6 nm | 1.6 nm | 0.8 nm | 0.8 nm | 0.8 nm |

**Table T1: Input parameters for all considered atomistic models.**

| Characteristic | Model A | Model B | Model C | Model D | Model E |
| --- | --- | --- | --- | --- | --- |
| Graphene sheet length (L) | 2.6 nm | 2.9 nm | 2.4 nm | 2.1 nm | 2.5 nm |
| Wavelength (λ) | 5.0 nm | 3.4 nm | 3.0 nm | 4.3 nm | 7.0 nm |
| Height of ripple | 0.15 – 0.25 nm | 0.15 – 0.25 nm | 0.25 -0.3 nm | 0.25 -0.3 nm | 0.35-0.5 nm |
| Rippling angle [˚] | 166(±1°) | 169(±1°) | 172(±2°) | 170(±2°) | 171(±3°) |
| Average rippling angle [˚] | | | 170°(±2°) | | |

**Table T2: Summary of the simulation results for all considered atomistic models.**

To test a possible effect of chirality we also performed simulations with the graphene sheet rotated. The final snapshot for the Model B with two different orientations of the graphene are indicated in Fig. S2 (a-b). In Fig. S2(c) the test strain simulations of the graphene ribbon (20nm long and 5nm wide) suspended in a vacuum are presented, indicating that mechanical response up to 10% strain should not significantly depend on the chirality of the considered graphene sheet.

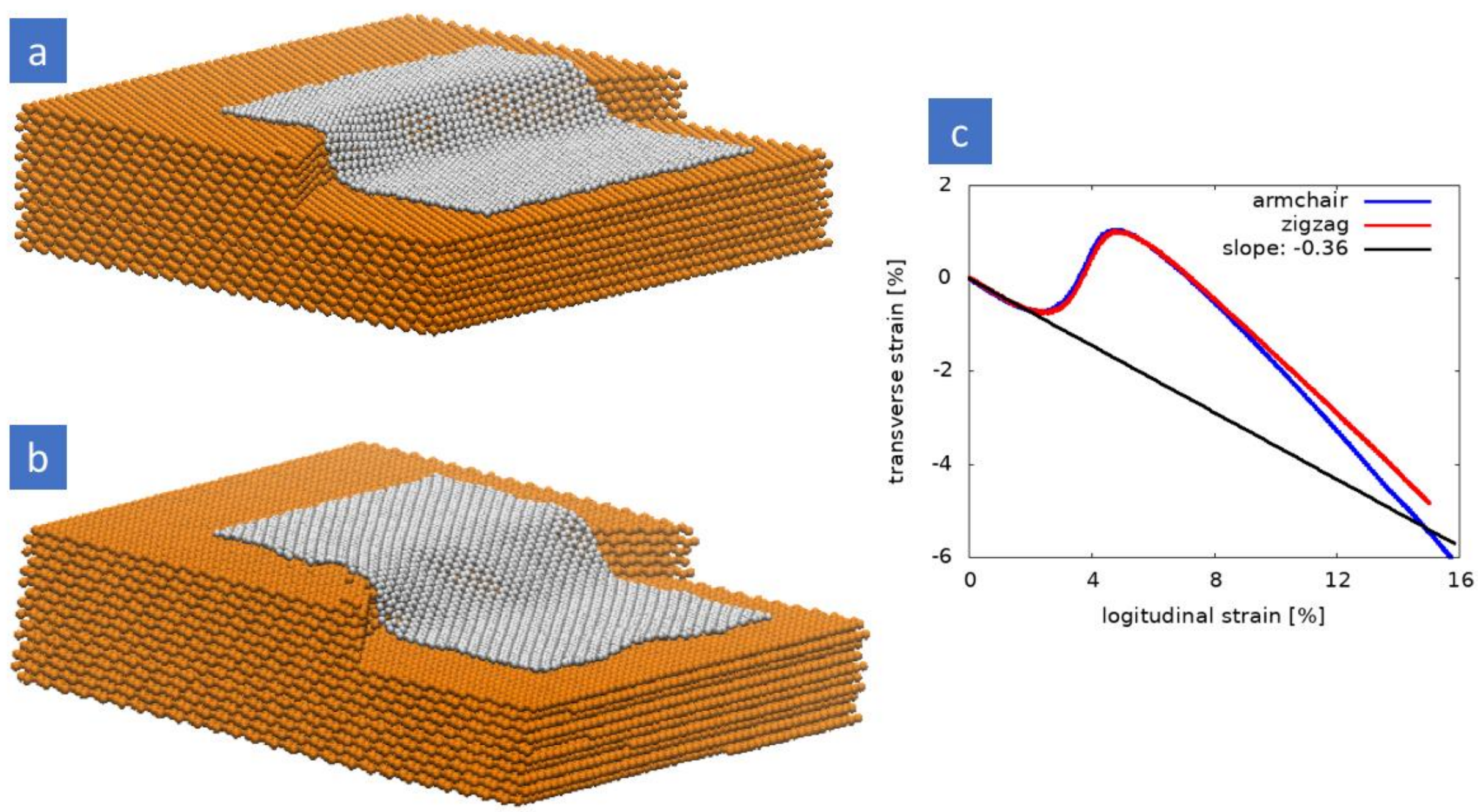

**Figure S2: Simulation of draped graphene sheet with different orientations:** (a) zigzag or (b) armchair direction along the copper step edge. (c) The strain simulations of the graphene ribbon (20nm long and 5nm wide) suspended in a vacuum at 300K with the shorter edge (sides lying on flat terraces) being either armchair or zigzag.

## 4. Role of dihedral interaction

To identify the role of the dihedral term in forming a ripple in a graphene sheet suspended over a curved Cu step edge, we repeated the ReaxFF simulation for model A in Table T1 with all dihedral interactions between the carbon atoms set to zero. The final structure for this simulation is presented in Fig. S3. Clearly, the removal of the dihedral interaction prohibits ripple formation. Such an observation implies that the ripple formations crucially depend on dihedral interaction between nearby carbon atoms.

Next, we study the spatial range of the dihedral interaction. For this, we performed a series of bond scan simulations, moving only one carbon atom out of a graphene plane. A small graphene sheet with hydrogenated edges was first energy minimized. Then, to achieve the out-of-the-plane displacement, bond distances with three adjacent C atoms were increased simultaneously by the same amount. This forces the C atom to move out of the plane perpendicular to the 2-D plane of the graphene sheet. The schematics of the considered cases and results are presented in Fig. S4. We considered eight scenarios, where the number the atoms that were allowed to relax during the simulation varies. The carbon atom that was moved out of plane relative to the graphene sheet is indicated in red in Fig. S4; all carbon atoms that were constrained to stay in plane are indicated in black; while the atoms that were allowed to relax are indicated in gray. We performed eight such simulations, where the range of the neighboring atoms allowed to relax, given by N, varies from 1 to 8. The changes in the potential energies for these eight systems compared to the case were only the nearest neighbors can relax are presented in Fig. S4. The energy difference is the greatest for the first nearest neighbors, implying the dihedral interactions are the strongest for them, whereas the energy difference between the two cases where 7 and 8 nearest neighbors are allowed to relax are almost the same, i.e., the effect of displacing the central C atom persists up to around 7 nearest neighbors. This sets the range of the in-plane dihedral interactions. A range of $N \geq 7$ nearest neighbors correspond to a distance of approximately 0.8 nm. In supplementary section 5 we measure the bending radius at the crest of a ripple and find it to be comparable to 0.8 nm. This observation adds further credence to the crucial role played by dihedral interactions in determining the shape of ripples.

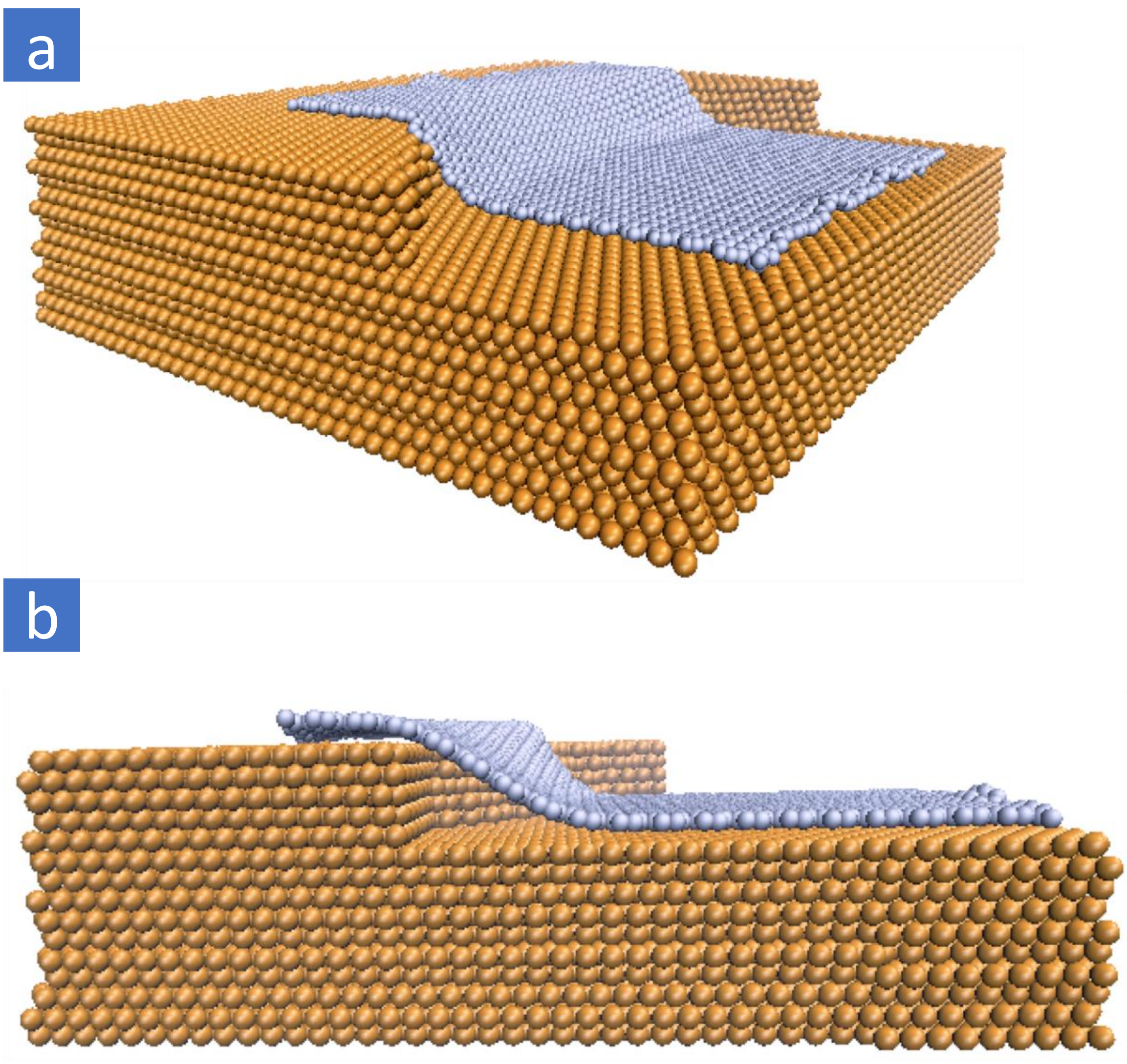

**Figure S3: Simulating draped graphene without dihedral interactions:** (a) Tilted and (b) split views show that ripples do not form in draped graphene sheets for the simulations with modified ReaxFF parameters, specifically, when there is no dihedral interactions between carbon atoms (in contrast to Fig. 4c and d of main text).

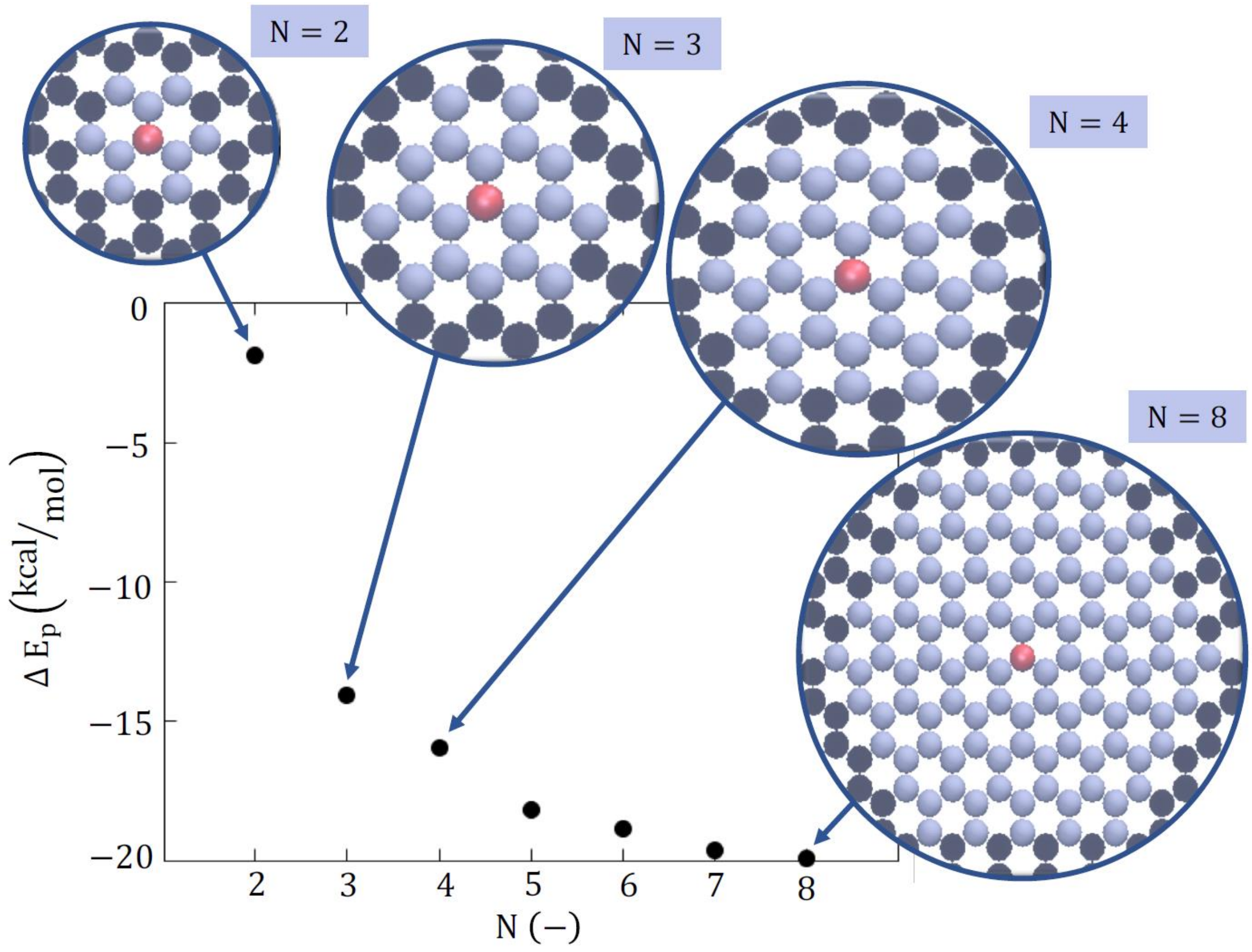

**Figure S4: Studying range of dihedral interactions:** By displacing one carbon atom out of plane (red), we see how far the effect persists in the graphene lattice by allowing more atoms around it to relax (grey). Atoms colored in black are kept fixed for the simulation. On the graph, the differences in the potential energy for the cases when a given range of the neighboring atoms are allowed to move compared to the case when only three nearest neighbors of the displaced atom are allowed to move are presented.

## 5. A phenomenological model for graphene bending

In the main text, we discussed how a dihedral interaction between neighboring $p_z$ orbitals can force graphene to stretch (Fig. 5e) and violate the classical theory of rippling. With this potential shortcoming in the classical theory of rippling identified using atomistic calculations, in this supplementary section we show that the shape of draped, rippled graphene can be explained with a simple phenomenological model. In particular, we modify the energy configurations we presented earlier for a classical sheet to include an extra "stretching due to bending" energy contribution $U_{sb}$:

$$U_{\mathrm{b}} = t^3 \int E(curvature)^2 \, dx \, dy \rightarrow t^3 \int E(curvature)^2 \, dx \, dy + t \int E \, \gamma_{sb}^2 \, dx \, dy = U_{\mathrm{b}} + U_{\mathrm{sb}}$$

This new energy contribution ($U_{sb}$) results in a transverse stretching strain $\gamma_{sb}$ over the bent region. If graphene bends with a radius of curvature $R$ at the crests and troughs (Fig. S5b), with the bent region (colored yellow in Fig. S5b) subtending an angle $\varphi$ at the center of the circle, then the bending energy is $U_b \sim Et^3 \int \left(\frac{1}{R^2}\right) dx\, dy \sim \frac{Et^3 L\varphi}{R}$, as the bending region has an arclength of $R\varphi$. The resultant stretching energy coming from this bending is $U_{sb} \sim Et\, {\gamma_{sb}}^2 (R\varphi) L$, where for simplicity we have assumed $\gamma_{sb}$ to be spatially uniform in the stretched region. The energy cost due to bending $U_b$ tries to increase the radius of curvature, but the associated stretching term $U_{sb}$ penalizes bending with large radii of curvature. The final radius of curvature is then determined by the balance of these two competing energy scales, where graphene is strain-locked.

$$\frac{d(U_b+U_{sb})}{dR} = 0 \Rightarrow R = \frac{t}{\gamma_{sb}} \quad \text{(Eq. S1)}$$

This simple phenomenological model can well explain our observations. First, having an optimal bending radius means graphene should always bend the same way, irrespective of the boundary conditions, as we have observed. While equation S1 implies the existence of an optimal bending radius, an accurate value of *R* is difficult to calculate from it as it involves the effective thickness *t* of the graphene sheet, estimation of which varies widely in the literature [8]. Instead, we measure the bending radius directly (in supplementary section 5) to be $R = 0.8\ \mathrm{nm}$. Most of the in-plane stretching of bonds should be limited to nearest neighbors as the $p_z$-$p_z$ dihedral interaction is strongest for them (see supplementary section 3). Thus, with the arclength of the bent region to be the C-C bond length (0.142 nm), we get $\varphi = \frac{0.142\ nm}{0.8\ nm} = 0.2\ \mathrm{rad} = 11°$. The calculated rippling angle $\theta = 180° - \varphi = 169°$ is consistent with our observed value of 168°($\pm$3°) (in Fig. 3d). Second, this consistency suggests that the bending and transverse stretching of graphene is limited to a very small region - of the order of graphene unit cell dimensions – instead of being distributed as a smaller stretch across all bonds. Thus, we expect triangular ripples, as observed. Finally, from equation S1, the bending radius *R* is also independent of the sheet length *L*, which explains why, contrary to the classical prediction (equation 2 of main text) there is no relation between wavelength $\lambda$ and *L*. Neither the wavelength - which depends only on the local step edge curvature (Fig. 2c, d and table T1) and hence can vary independently - nor the fixed bending radius *R* affect the effective sheet length *L*.

In the above analysis, we assumed that the contribution of the longitudinal stretching energy $U_{st}$ is negligible over the narrow bending region compared to the bending ($U_b$) and transverse stretching ($U_{sb}$) energies. Next, we justify this assumption. In particular, we argue $U_{st} \ll (U_b + U_{sb})|_{R = t/\gamma_{sb}}$, which implies that longitudinal stretching plays a minor role in determining the ultimate shape of the ripples. Analogous to the classical case (derived in supplementary section 9), the contribution of the longitudinal stretching energy integrated over the bent region is $U_{st} = \int T \left(\frac{d\xi}{dx}\right)^2 dx\, dy = Et\gamma \left(\frac{A}{L}\right)^2 L(R\varphi)$. The ratio $\frac{U_{st}}{U_b+U_{sb}} = \frac{Et\gamma A^2 R\varphi/L}{2Et^3 L\varphi/R} = \frac{\gamma A^2 R^2}{2t^2 L^2}$. We can make an order of magnitude estimate of this ratio from our experiment with extreme strain $\gamma \approx 0.1$, $R \approx 0.8\ \mathrm{nm}$ (measured directly in supplementary sec. 5) and $L \approx 20\ \mathrm{nm}$ (from Fig. 3e). Assuming the thickness to be the interlayer spacing of graphite [8,9] would make $t = 0.335\ nm$. On the other hand, ab-initio bond-orbital models [10] estimate the effective thickness to be as low as 0.08 nm. As graphene bends with a conserved rippling angle of 168° at the crest (Fig. 3d), the amplitude *A* is related to the wavelength λ by the relation $A = \frac{\lambda}{2\, tan(168°/2)} \simeq \frac{\lambda}{20}$. Even considering the smaller thickness $t = 0.08$ nm and wavelengths of $\lambda \approx 25$ nm (from Fig. 3e), $\frac{U_{st}}{U_b+U_{sb}} = \frac{\gamma\lambda^2R^2}{800(t^2L^2)} = 0.02 \ll 1$, proving that the energy contribution from longitudinal stretching $U_{st}$ is indeed negligible over the narrow region over which graphene bends.

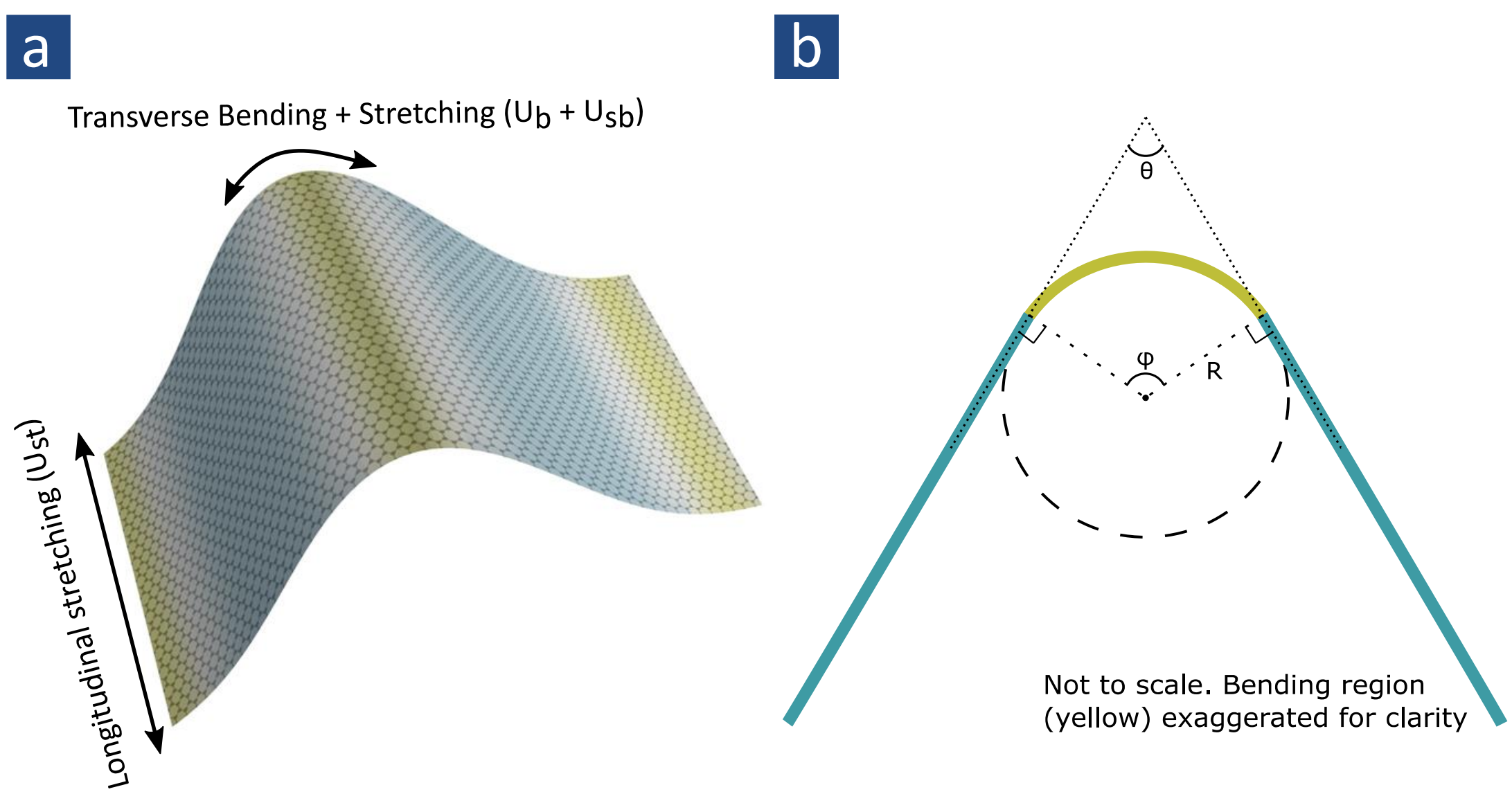

**Figure S5: A phenomenological model for graphene bending:** (a) Analogous to a classical fabric under tension in the longitudinal direction, graphene ripples in the transverse direction. However, unlike a classical fabric which does not stretch significantly in the transverse direction when bending, graphene undergoes significant stretching when bent, resulting in triangular ripples, where bending is confined to a narrow region with dimension on the order of a few unit cells. The crests and troughs of the ripples (yellow, where the graphene bends) undergo significant stretching in the transverse direction compared to the unbent regions (turquoise). As observed in Fig. 3d and explained using a phenomenological model, balance of the transverse bending and stretching energies ($U_b$ *and* $U_{sb}$ *respectively*) cause graphene to always bend with the same angle (θ in (b), and measured in the experiment in Fig. 3d). The effect of the longitudinal stretching energy $U_{st}$ over the narrow bending region is negligible, as explained above. (b) Schematic side view of (a) showing graphene bending with a radius of curvature R. Features are not to scale. The bending region (yellow) is exaggerated for clarity resulting in a rippling angle θ at the vertex that is much smaller than our observed 168°.

## 6. Measuring the radius of curvature of graphene at ripple crests

Measuring an accurate value of the bending radius of graphene is required to calculate the rippling angle from equation S1. As argued in our phenomenological model, transverse in-plane stretching of graphene results in an optimal bending radius at the crest of each ripple. However, equation S1 could not be used to estimate this radius as it requires estimating the thickness of graphene, a value which varies widely in the literature [8]. Instead, we measured the optimal bending radius directly from our data.

To estimate the bending radius, we take high resolution images of ripples and extract line cut profiles from them (Fig. S6). Note that unlike the previous depiction (Fig. 3c), the horizontal and vertical scales in Fig. S5 are equal. The points at the crests of the ripple are fit to a circle using an optimization routine in the scipy package [11]. In particular, we use the optimization routine to fit a circle such that the sum of squared deviations of the fitted circle compared to the data is minimized. The fitting yields a bending radius of curvature of 0.8 nm. We note that this length scale of 0.8 nm is the approximate length of the dihedral interaction important in ripple formation (Fig. S4).

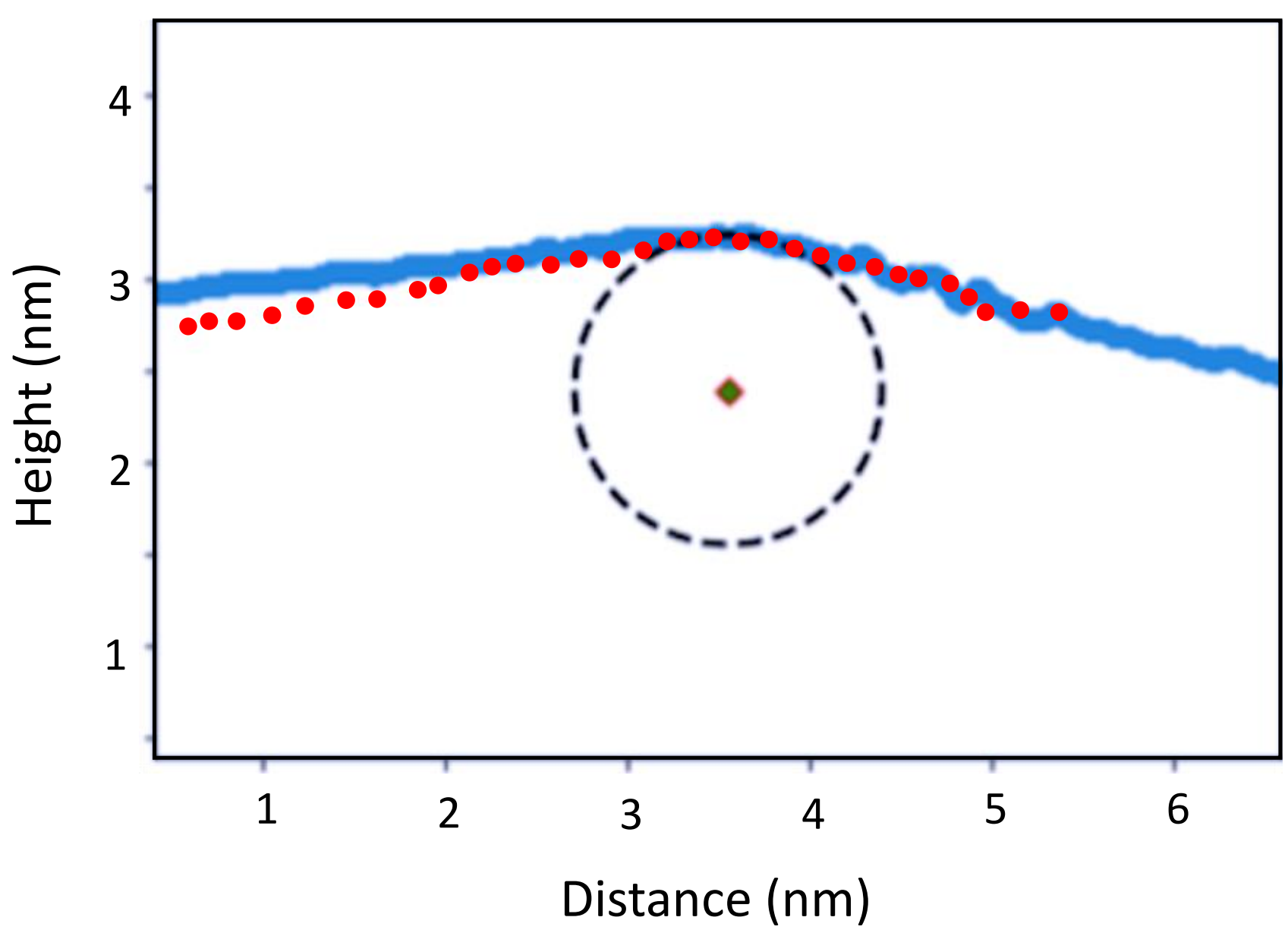

**Figure S6: Measuring the radius of curvature at the ripple crest by fitting a circle.** The blue curve is a high resolution ripple profile. The best fit circle (black dashed line) has a radius of R = 0.8 nm. Note that unlike in Fig. 3c, the horizontal and the vertical axes here have the same scale. The red dots overlaid on the experimental data are obtained from a linecut profile of a ripple from our ReaxFF simulations (model A in table T1).

## 7. Possible artifacts in imaging a tilted surface

Generally, STM measurements portray a well approximated description of a sample surface down to atomic level corrugations. However, careful consideration of potential scanning artifacts is necessary for cases where the point of contact between the tip and sample can change during a scan, like when imaging regions with high slopes or large corrugations. In such cases, the finite size of the tip can create artifacts that need to be considered for proper interpretation of the data. Such a scenario indeed arises in our case when scanning the draped region, and it is imperative to address if any of our observations are consequences of scanning artifacts instead of actual topographic features.

Fig. S7a shows the imaged height profile of a draped region. Note the rounded upper edge but sharp bottom one. In Fig. S7b, we argue that this feature is an artifact of having a tip of finite size. As the tip scans across a step edge, the point of contact between the tip and surface changes. The rounded top edge results from the tip scanning itself at the top corner. On reaching the bottom terrace, however, the point of contact between the tip and sample is abruptly changed, resulting in a sharp kink. Thus, the actual curvature of the graphene sheet at its top and bottom corners cannot be imaged by a finite sized tip.

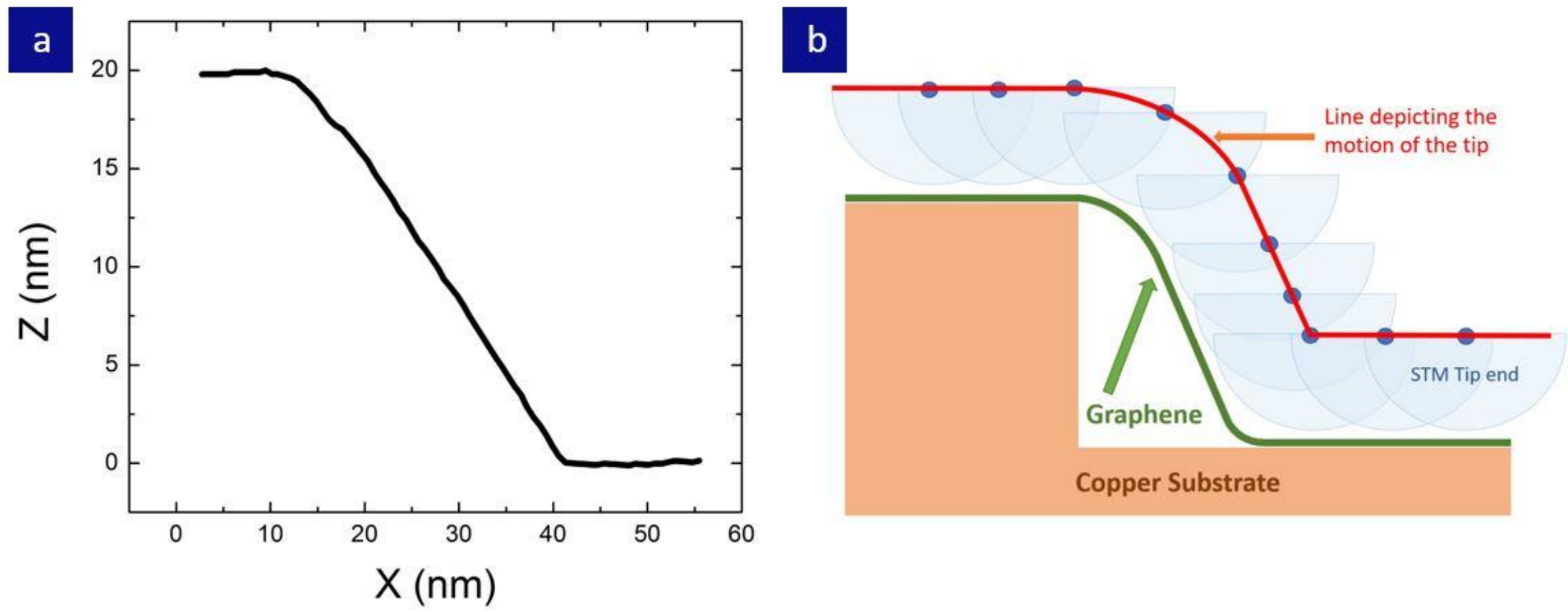

**Figure S7: Tip artifacts at top and bottom of draped region** (a) Height profile of draped graphene shows a rounded edge at the top and a sharp edge at the bottom. (b) This is an artifact of scanning with a tip of finite size, as the contact point of the tip with the sample changes in the draped region. See text of supplementary section 6 for explanations.

When measuring the height profile of an atomically corrugated surface using a scanned probe technique, the sample rarely lies exactly in the XY scan plane. Small tilts are usually removed using plane subtraction. However, when the tilt is large (as in our draped graphene), this typical STM analysis approach of using plane subtraction leads to artificial compression along the draping direction (it is equivalent to projection into a plane parallel to the terraces – see Fig. S6). Thus, we instead locally rotate the coordinate system in order to properly extract all distances in the suspended material. In order to do this we begin with the global ($\hat{u}, \hat{v}, \hat{w}$) coordinates of every pixel in a field of view. We identify the top and bottom edge of a draped region by looking for a sharp change in slope. At every point on the draped region we are then able to define local unit vectors ($\hat{x}, \hat{y}, \hat{z}$) as follows: we define the local x-axis along lines parallel to the edges, allowing a slight tilt to match the steepest slope (that is, the ripple direction), we define the y-axis as perpendicular to this axis, connecting the two edges, and finally we filter out the ripples along the local x-axis (i.e. along one-dimensional lines parallel to the ripple edges), allowing us to define the local z-axis as perpendicular to this ripple-removed sheet, and redefine the x-axis by projecting it into this sheet. Once we have defined local coordinates everywhere in the ripple-removed sheet we assign (x, y, z) coordinate values by moving from pixel to pixel and projecting the displacements onto the local axes. This method of defining local coordinates allows us to “unwrap” the complex draped, curved and rippled sheet into an essentially rectangular band of z-values without the artificial skewing that would accompany traditional plane subtraction (See Figure S8).

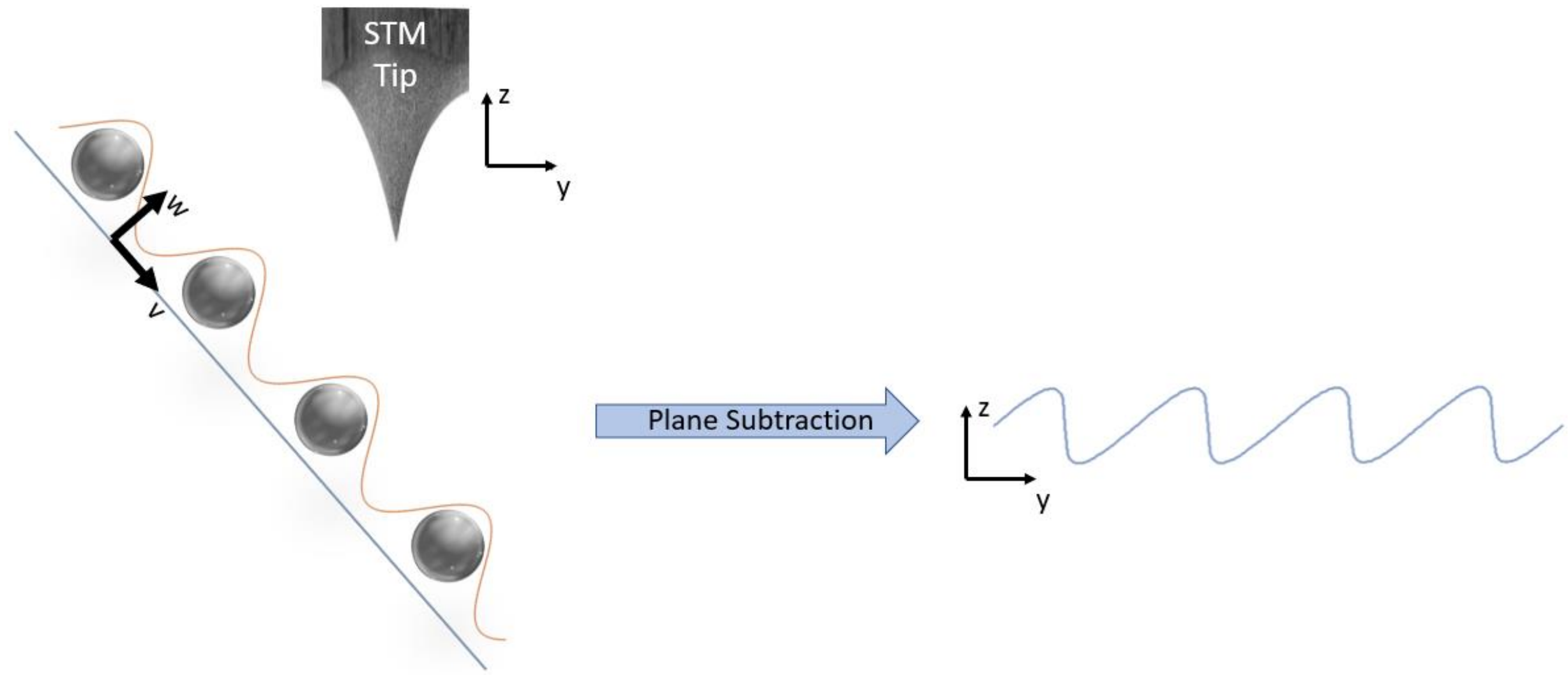

**Figure S8**: **Plane-subtraction on a tilted surface.** When imaging the height profile of an atomically corrugated surface (atoms represented by grey spheres), the STM tip moving along the z-height and y-direction follows the red curve. Using the standard analysis technique of plane subtraction results in a profile that is artificially skewed (blue curve on the right). To extract the correct surface features we work in a local coordinate system defined on the draped region (u,v,w) instead of the (x,y,z) coordinate system of the tip. The details of transforming to the (u,v,w) coordinate system from the (x,y,z) coordinate system (in which the data is originally collected) is provided in supplementary section 6. The tilt and skewing are exaggerated for clarity.

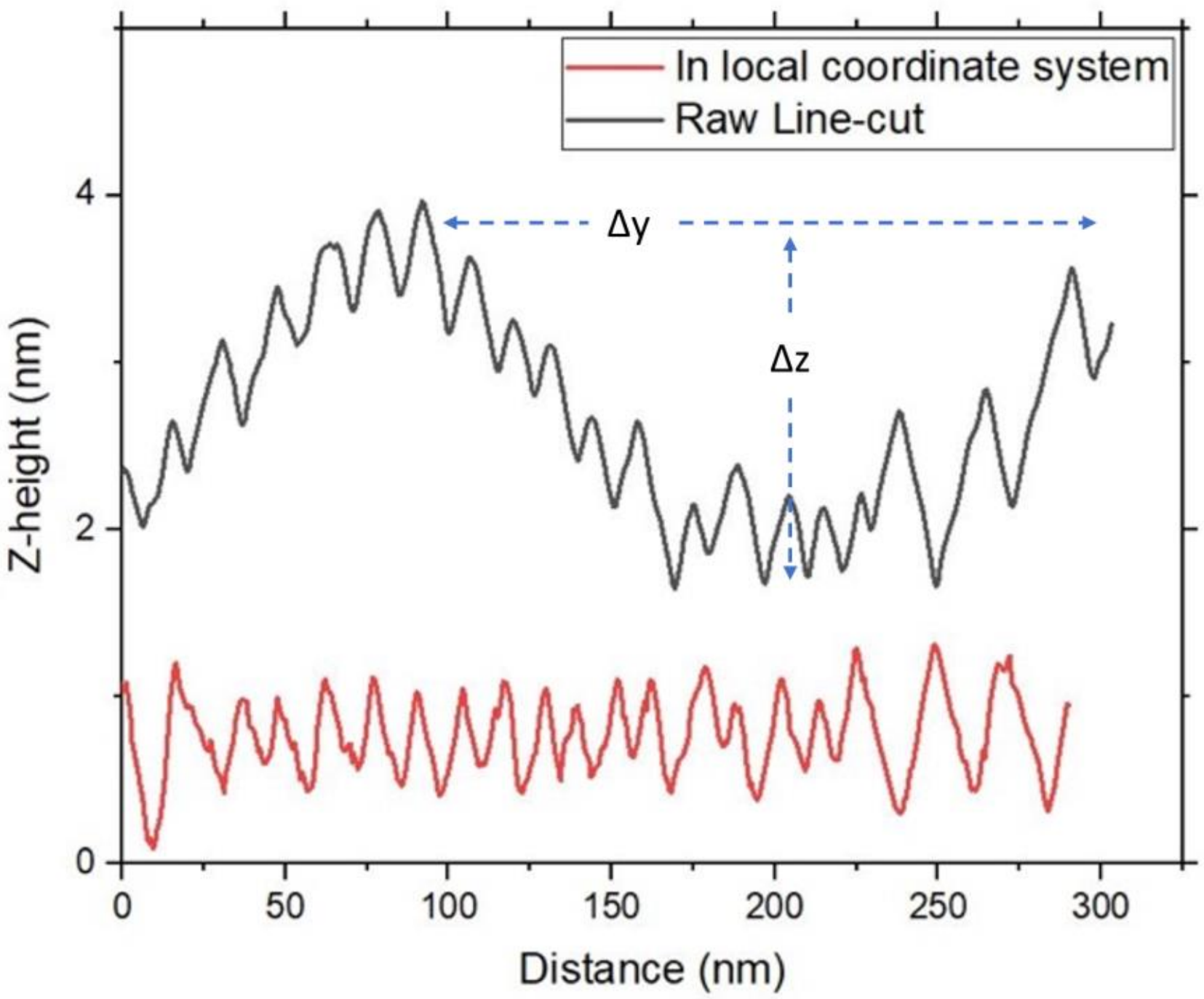

**Figure S9**: **Transformation to local coordinate system:** While the ripples in the global coordinate system (black) are skewed, the skewing is significantly reduced in the local coordinate system (red) of the ripples. From line cuts like the black curve, we can make also make an order of magnitude estimation of the radius of curvature of the step edges underneath the ripples. As curvature $\approx \frac{2\Delta z}{\Delta y^2} \approx \frac{4\ nm}{(200\ nm)^2}$, we get $R_{step} \approx 10\ \mu m.$

## 8. Transverse stretching at ripple extrema from simulations

Density functional theory calculations of a rolled up carbon nanotube showed that its radius increases due to dihedral interactions [12]. In the main text of the paper (Fig. 5), we saw signatures of this transverse stretching in the local density of states as the graphene sheet is bent. In Fig. S10 we measure the bond lengths in the simulated system and see the transverse stretching at the ripple crest. Though ReaxFF does not account for electronic interactions explicitly, the general trend of longer bond lengths is seen near the ripple crest. However, we note that the magnitude of strain variation at and away from the crest is an order of magnitude less than what was estimated experimentally by measuring the Dirac point shift. This might be attributed to the current limitations in the force field. While the current force field parameters are sufficient to describe the mechanical properties and reproduce the rippling angle, an eReaxFF calculation, which can explicitly account for orbital adjustments may be able to capture the spatial strain variations better. Also important will be accurate modeling of the growth process, especially the diffusive motion of the Cu step edges which can increase longitudinal strain in the ripples. This extra longitudinal strain can affect the transverse strain (due to Poisson effect) and also help stabilize the ripples formed to reduce the large error bars in Fig. S10d. Step edge diffusion is slower compared to the time required for ripples to form, and would require simulating much longer time scales than what is considered in the present ReaxFF calculations.

ReaxFF also cannot directly determine the shape of molecular orbitals. To visualize the molecular orbitals for bent graphene and any possible changes in the shape of these molecular orbitals, DFT calculations using the Schrodinger software were performed [13]. A carbon nanotube (CNT) was used, as it is the extreme case of graphene bending. First, the CNT was hydrogenated at the edges and then the structure was energy minimized using the B3LYP hybrid functional with a 6-31g+ basis set. The energy minimized structures were then used to calculate the molecular orbitals. The final molecular structures along with the orbitals were plotted using the Jaguar software [14] with the default isosurface cutoff value 0.02, to get Fig. 5d and e.

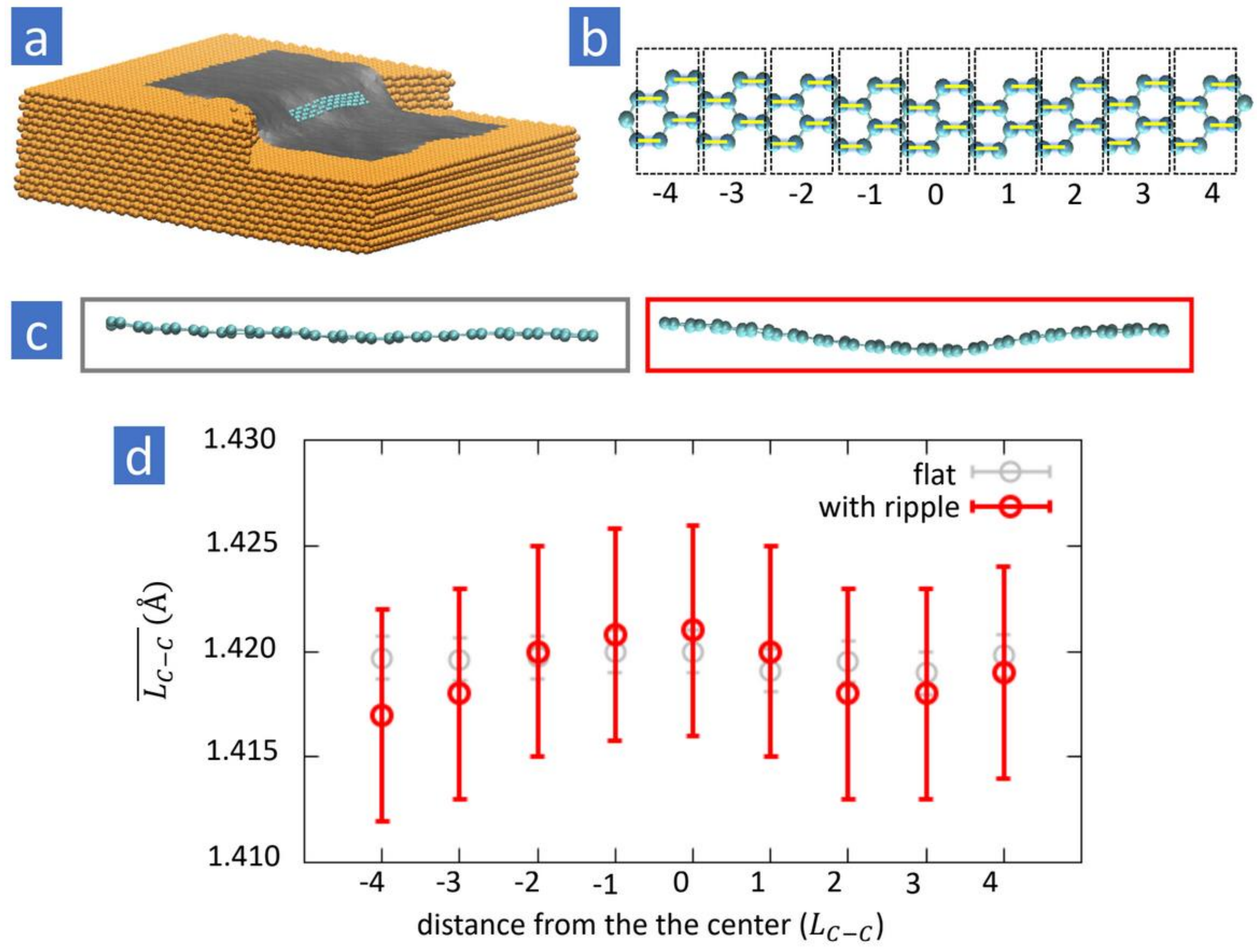

**Figure S10**: **Transverse stretching from ReaxFF:** Considering three lines of atoms across the ripple (a), marked by cyan lines, we break them into regions near the ripple crest (b). The average bond length is calculated in each box (numbered -4 to 4) and plotted in (d). The average bond length in rippled graphene varies spatially (red), peaking at the ripple crest, whereas the bond lengths of a flat graphene sheet (grey) do not change significantly.

## 9. Scaling laws for draped fabrics

In the main text of the paper, we presented a scaling law for classical fabrics stretched under uniform strain, $L \propto \lambda^2$, and used it to demonstrate that graphene does not behave like a classical fabric. In this supplementary section, we sketch the origin of this scaling law. Our analysis closely follows that of Cerda et. al. in ref. [15], where they derived it for a classical fabric stretching in a flat geometry and ref. [16], where they studied possible modifications to it when it is draped over a curved table.

First, we note that though the underlying step edge profile is not imaged directly, we can make an order of magnitude estimate of its curvature by taking line cuts as shown in Fig. S9, where a typical profile of ripples adorning a step edge is shown, and from whence the line cut profiles in Fig. 3 of the main text are extracted. Such a profile demonstrates that the radius of curvature of the Cu step edges, $R_{step}$, is typically several orders of magnitude larger than the other length scales in the problem ($R_{step} \gg A, \lambda \text{ and } L$). For example, for the profile shown in Fig. S9, the ripple amplitude *A* is of the order of a few nanometers, the wavelength $\lambda$ and sheet length *L* are of the order of tens of nanometers, and $R_{step} \approx 10\ \mu m$. This situation is similar to the commonly observed ripples in a tablecloth adorning a round table, where the radius of curvature of the table is much larger than the ripple wavelength, and is also one of the cases considered in ref. [16].

As a sheet under tension in the longitudinal direction ripples in the transverse direction as, the shape of ripples is determined by balancing the energy contributions from these two deformations (Fig. 1a of main text). If we consider a sheet of thickness *t*, length *L* and Young's modulus *E*, we find that rippling under applied tension per unit length *T* results in a longitudinal strain $\gamma = \frac{T}{Et}$. The energy due to this longitudinal extension, integrated over a wavelength $\lambda$ of the ripples, is $U_{\mathrm{st}} \sim \int T \left(\frac{d\xi}{dx}\right)^2 dx\, dy$, analogous to the energy stored in a stretched string, where $\xi$ is the out of plane displacement. This leads to an energy associated with longitudinal stretching

$$U_{st} \sim Et\gamma \left(\frac{A}{L}\right)^2 L \cdot \lambda = \frac{Et\gamma A^2 \lambda}{L}, \qquad \text{(Eq. S2)}$$

where *A* is the ripple amplitude (z-distance from ripples' crest to base). On the other hand, the bending energy of a ripple is $U_b \sim t^3 \int E(curvature)^2\, dx\, dy$. The curvature of the sheet as it ripples is approximately $\frac{A}{\lambda^2}$, and the radius of curvature of the underlying step is $R_{step}$ (absolute value), making the total bending energy integrated over a ripple wavelength to be

$$U_b \sim Et^3 \left(\frac{A}{\lambda^2} + \frac{1}{R_{step}}\right)^2 L\lambda \approx Et^3 \left(\frac{A^2}{\lambda^4} + \frac{2A}{R_{step}\lambda^2}\right) L\lambda \qquad \text{(Eq. S3)}$$

where we have dropped terms proportional to $\frac{1}{R_{step}{}^2}$. Equations S2 & S3 imply that $U_{st}$ tries to reduce the wavelength $\lambda$ while $U_b$ tries to increase it. The ultimate shape of the ripples results from a balance between these two energies, and so, $U_{st} \sim U_b$ yields the scaling law

$$\gamma \sim t^2 L^2 \left(\frac{1}{\lambda^4} + \frac{2}{R_{step} A \lambda^2}\right) \qquad \text{(Eq. S4)}$$

Considering a ripple profile $\xi(x, y)$ extracted at a given height on a curved step edge, if $\lambda'$ is the length of fabric forming a ripple of wavelength $\lambda$, then, the inextensibility of the thin classical fabric implies

$$\lambda' = \int_0^{\lambda} \sqrt{1 + \left(\frac{\partial \xi}{\partial y}\right)^2}\, dy = \lambda + \frac{1}{2} \int_0^{\lambda} \left(\frac{\partial \xi}{\partial y}\right)^2 dy \qquad \text{(Eq. S5)}$$

using the binomial approximation. As the second term on the right is the 'extra length' $\Delta\lambda$ that is rippling, we must have $\Delta\lambda = \frac{1}{2}\int_0^{\lambda}\left(\frac{\partial\xi}{\partial y}\right)^2 dy$. Using an ansatz $\xi = \xi_0(y) + A\cos\left(\frac{2\pi y}{\lambda}\right)$, where $\xi_0(y)$ is the curved step edge profile given by $\xi_0(y) = \sqrt{{R_{step}}^2 - y^2}$, we have

$$\Delta\lambda = \int_0^{\lambda}\left[\frac{y^2}{R_{step}^2 - y^2} - \frac{2\pi A^2}{\lambda^2}\sin^2\left(\frac{2\pi y}{\lambda}\right) + \frac{4\pi y\, A\sin\left(\frac{2\pi y}{\lambda}\right)}{\lambda\sqrt{R_{step}^2 - y^2}}\right]dy \sim \frac{2A^2}{\lambda} + \frac{4A\lambda}{\sqrt{R_{step}^2 - \lambda^2}} \quad \text{(Eq. S6)}$$

where we have again dropped terms proportional to $\frac{1}{{R_{step}}^2}$ on the right hand side. Using ${R_{step}}^2 - \lambda^2 \approx {R_{step}}^2$, we can simplify it to get

$$\nu\gamma = \frac{\Delta\lambda}{\lambda} \sim \frac{2A^2}{\lambda^2} + \frac{4\,A}{R_{step}} \quad \text{(Eq. S7)}$$

Equating $\gamma$ from equations S4 and S7 and replacing $A$ by $\lambda$ as all ripples we observe have a conserved vertex angle (Fig. 3d of main text), meaning that the two are not independent variables and related by $A \sim \lambda$, we get

$$\lambda \sim R_{step} \quad \text{(Eq. S8)}$$

Equation S8 implies that up to first order (as we have neglected all terms proportional to $\frac{1}{{R_{step}}^2}$ ), the wavelength and the radius of curvature of the underlying step edge are proportional. We note that such a relation is to be expected, because if there are $n$ ripples around a cylindrical table, the wavelength should satisfy $n\lambda \approx 2\pi R_{step}$, as also noted in ref. [16].

With $\lambda \sim R_{step}, A$, from equation S4, we recover the same scaling law presented in Equation 2 of the main text, relating $\lambda$ and $L$:

$$L \sim \frac{\sqrt{\gamma}\lambda^2}{t}$$

The above analysis demonstrates that up to first order, while the local wavelength depends on the local radius of curvature of the step edge, it does not change the scaling law $L \propto \lambda^2$ for classical fabrics. Any higher order correction term to equation S4 should be at least of order $O(\frac{L^2}{{R_{step}}^2})$ or $O(\frac{\lambda^2}{{R_{step}}^2})$, which would be of the order of $10^{-6}$, as $R_{step} \approx 10^3\, L, \lambda$. Equation S8 implies that small local changes in $R_{step}$ can change the local wavelength. However, as all observed ripple wavelengths are within an order of magnitude (Fig. 3d of main text), this implies that classically, $R_{step}$ can only vary within an order of magnitude, which is still much larger than all other length scales in the problem, meaning the above argument still holds and local variations in $R_{step}$ cannot account for any claimed non-classical behavior.

Referring to the work of ref. [16], we note that there a thick sheet with large mass was considered. Thus, the tension in the sheet increases as a function of sheet length $L$ (because of gravity), resulting in the stretching energy to be $U_{\mathrm{st}} \sim \int T\left(\frac{d\xi}{dx}\right)^2 dx\,dy \sim L\left(\frac{A}{L}\right)^2 L \cdot \lambda \sim A^2\,\lambda$. With the bending energy remaining unchanged from what derived in equation S3 above, $U_{\mathrm{st}} \sim U_{\mathrm{b}}$ yields a scaling law of $A^2\,\lambda \sim \frac{A^2 L}{\lambda^3} \Rightarrow L \sim \lambda^4$, as was derived in ref. [16] (near bottom of page 4 of their paper). But in our case, the flimsy graphene sheet is unaffected by gravity and the tension is determined by the van der Waals forces attaching it to the Cu substrate, and so does not vary along the sheet length, implying that the scaling law $L \propto \lambda^2$ should hold in our case if graphene behaved like a classical sheet.

The above analysis demonstrates that the scaling law $L \sim \sqrt{\gamma}\lambda^2$ presented in the main text should have very small corrections for the gently curved step edges observed experimentally. However, this curvature, small as it is, seems to be a crucial component in the formation of ripples in graphene, as demonstrated in our ReaxFF calculations (Fig. 4). As graphene behaves like a non classical fabric, the scaling laws derived in this supplementary section should be modified for nanoscale ripples observed in graphene. In particular, local stretching of the bonds at the ripple crests (demonstrated by Dirac point shift in Fig. 5c) implies that the inextensibility condition in equation S5 is violated. We hope to theoretically derive and experimentally demonstrate the scaling laws for such a non classical fabric in a future publication.